\documentclass[fleqn,usenatbib,useAMS]{mnras}

\usepackage{graphicx}	% Including figure files
\usepackage{amsmath}	% Advanced maths commands
\usepackage{amssymb}	% Extra maths symbols
\usepackage{multicol}      	% Multi-column entries in tables
\usepackage{bm}		% Bold maths symbols, including upright Greek
\usepackage{pdflscape}	% Landscape pages
\usepackage{xcolor}		% more colors

\DeclareRobustCommand{\VAN}[3]{#2}
\let\VANthebibliography\thebibliography
\def\thebibliography{\DeclareRobustCommand{\VAN}[3]{##3}\VANthebibliography}

\usepackage[T1]{fontenc}
\usepackage{ae,aecompl}

\usepackage{newtxtext,newtxmath}

\title[Missing Metals in DQ Stars]{Missing Metals in DQ Stars; a Compelling Clue to their Origin\thanks{Dedicated to the memory of Prof. G. Fontaine (1948--2019)}}

\author[J. Farihi et al.]{J. Farihi$^1$\thanks{E-mail: j.farihi@ucl.ac.uk}, P. Dufour$^2$, T. G. Wilson$^{1,3,4}$
\\
$^1$Department of Physics and Astronomy, University College London, London WC1E 6BT, UK\\
$^2$D\'epartement de Physique, Universit\'e de Montr\'eal, Montr\'eal, Qu\'ebec H3C 3J7, Canada\\
$^3$Isaac Newton Group of Telescopes, E-38700 Santa Cruz de La Palma, Spain\\
$^4$School of Physics and Astronomy, University of St. Andrews, St. Andrews KY16 9SS, UK
}

\begin{document}

%\label{firstpage}

\maketitle

\begin{abstract}
White dwarf stars frequently experience external pollution by heavy elements, and yet the intrinsically carbon-enriched DQ spectral class members fail to exhibit this phenomenon, representing a decades-old conundrum.  This study reports a high-resolution spectroscopic search for Ca\,{\sc ii} in classical DQ white dwarfs, finding that these stars are stunted both in pollution frequency and heavy element mass fractions, relative to the wider population.  Compared to other white dwarf spectral classes, the average external accretion rate is found to be at least three orders of magnitude lower in the DQ stars.  Several hypotheses are considered which need to simultaneously account for i) an apparent lack of accreted metals, ii) a dearth of circumstellar planetary material, iii) an observed deficit of unevolved companions in post-common envelope binaries, iv) relatively low helium mass fractions, and remnant masses that appear smaller than for other spectral classes, v) a high incidence of strong magnetism, and vi) modestly older disk kinematics.  Only one hypothesis is consistent with all these constraints, suggesting DQ white dwarfs are the progeny of binary evolution that altered both their stellar structures and their circumstellar environments.  A binary origin is already suspected for the warmer and more massive DQ stars, and is proposed here as an inclusive mechanism to expose core carbon material, in a potential evolutionary unification for the entire DQ spectral class.  In this picture, DQ stars are not descended from DA or DB white dwarfs that commonly host dynamically-active planetary systems.
\end{abstract}


\section{Introduction}

White dwarf stars whose spectra exhibit molecular carbon features have been known for over 60 years, when the first pair were discovered by carefully constructed spectrophotometry \citep{greenstein1957,bell1962}.  Owing to instrumentation limits on wavelength at that time, only the weaker C$_2$\,(1,0) feature near 4670\,\AA \ was observable \citep{eggen1965,greenstein1970}, and not the stronger C$_2$\,(0,0) feature with bandhead near 5165\,\AA \ that would later be found in many more stars (e.g.\ \citealt{oke1974,liebert1977,wesemael1993}).  It has been understood for half a century that white dwarfs exhibiting C$_2$ Swan bands (or other carbon features) in their spectra -- denoted as spectral class DQ \citep{mccook1987} -- have atmospheres that are dominated by helium, with only trace abundances of carbon, and little or no hydrogen \citep{bues1973,grenfell1974}.  

Early pioneering work to understand the carbon abundances in DQ stars pointed out that convective mixing in a gravitationally-stratified atmosphere was problematic, because full mixing would likely result, and thus lead to a carbon-dominated atmosphere by mass \citep{vauclair1979a,dantona1979}, in contrast with the most successful models for the observed spectra.  There were already theoretical indications that the mass of the helium layer in DQ stars should be significantly thinner than in their non-carbon bearing, helium atmosphere counterparts \citep{muchmore1977}, and this has been corroborated by successful models that dredge up carbon, where it emerges through the diffusive (upward) tail in a stratified atmosphere \citep{koester1982,fontaine1984}.  The first time-dependent modeling of this process showed that the inferred trace carbon abundances are best reproduced with helium layer mass fractions in the range $-4.0 \lesssim \log q({\rm He}) \lesssim -3.5$, and demonstrated that higher mass fractions, such as those predicted by stellar evolution, fail to dredge sufficient carbon to reproduce the observations \citep{pelletier1986}.

On a similar timescale, it was already known that white dwarfs sometimes exhibited metals such as Ca, Mg, and Fe in their spectra, and that these were also trace abundances in helium-dominated atmospheres (e.g.\ vMa\,2; \citealt{weidemann1960,strittmatter1971,wegner1972}.  In this way, the metal-rich DZ spectral class was shown to share basic characteristics, such as temperature and atmospheric composition, with the DQ stars, based on the earliest discoveries and successful modeling.  In contrast with the DQ stars, however, there was already a scientific consensus that the metals observed in DZ stars could not have been brought to the surface from their interiors \citep{fontaine1979,vauclair1979b}.  Over a wide range of temperatures, white dwarfs should be gravitationally sedimented (via diffusion), on timescales much faster than any potentially competing process \citep{dupuis1992}, including convective mixing -- a mechanism that actually enhances downward diffusion \citep{alcock1980}.  Thus it has been clear for many decades that an external source is necessary for DZ stars, but that DQ stars can be understood as polluted via their own interiors.

\citet{fontaine1984} first mentioned the striking problem that is the subject of this work, ``{\it... it is interesting to speculate on the reasons why metals and carbon still appear to be mutually exclusive in the atmospheres of cool helium-rich white dwarfs.}''  The DQ stars are around 9--10 per cent of the local population as defined by the 20\,pc sample, which is well constrained if not complete \citep{hollands2018}, and thus are not outliers or unusual by any means.  Indeed, more than 60 years after the first discovery, this fundamental question about their nature remains unanswered, despite recent and ongoing work with far superior data and modeling, and based on hundreds of DQ and DZ stars \citep{koester2019,coutu2019,koester2020}.  In response to an earlier version of the study presented here (Farihi et al.\ 2022; arXiv:2208.05990v1), it has been shown that it is possible to suppress or mask the Swan bands in stars with sufficient metal abundances resulting from external pollution, but not in all cases \citep{blouin2022}, and hence this conundrum remains.

This paper reports a deep spectroscopic search for external, heavy element pollution in DQ stars using one of the largest ground-based telescopes available.  A dramatic lack of metal pollution is revealed for DQ white dwarfs, as compared to the similarly helium-rich DZ stars.  Section~2 reports on the high-resolution spectroscopic observations, and Section~3 quantifies the differences between the DQ stars and DZ stars in terms of pollution frequencies and accretion rates, using both the high-resolution spectral data as well as available SDSS spectroscopy.  Section~4 examines other threads of available evidence on DQ stars, and in particular where there are distinct properties as compared to DZ-type and other white dwarfs (mass and stellar structure, circumstellar matter, duplicity, magnetism, and kinematics).  Section~5 examines several hypotheses that might account for all the available evidence, and concludes that the DQ white dwarfs are the result of binary evolution, where mass loss has resulted in a thinner helium layer than otherwise expected, and which also depleted any nearby planetary material that might pollute the star.  Section~6 discusses this possibility in detail, and provides suggestions for future work that can test this novel proposed origin for DQ stars.

%DQ stars missing are GJ\,86B and Procyon\,B, 2 of 11 stars not in DR2 but within 20\,pc
%In the nearly complete 20\,pc sample of white dwarfs \citep{hollands2018}, the DQ stars comprise 13/139 = 9.4 per cent (or 10.0 per cent if 2 of 11 missing (in DR2 and eDR3) 20\,pc members from their Table~4 are included).

\section{Spectroscopy and metal abundances}

 %%%TABLE TARGETS%%%
\begin{table}
\begin{center}
\caption{Target parameter summary, Ca upper limits and detections.\label{targs}}
\begin{tabular}{@{}crlcrl@{}}

\hline

WD\#		&$T_{\rm eff}$	&Mass		&[C/He]	&[Ca/He]	&Notes\\
			&(K)		&($M_{\odot}$)		&		&		&\\
				
\hline

0038$-$226   	&5210	&0.51		&$-$8.4	&$<-$12.3		&1\\		% CLINES:  	opt Swan bands, broad, shift DQp, bergeron+ 1994, my XS data
0042$-$238	&10500 	&0.5:			&$-$2.7	&$<-$11.8		&2\\		% CLINES: 	opt Swan bands, CI lines, koester+ 1982
0312$-$084	&9080  	&0.5:			&$-$4.2	&$<-$12.0		&3\\		% CLINES:	opt Swan bands, sayres+ 2012
0435$-$088	&6400  	&0.55		&$-$6.3	&$<-$12.8		&4\\		% CLINES:	opt Swan bands, dufour+ 2005
0548$-$001	&6080  	&0.66		&$-$6.6 	&$<-$12.5		&1\\		% CLINES:	opt Swan bands, CH band, dufour+ 2005, DQP = polarized, not DQp!
0806$-$661	&10210 	&0.58		&$-$5.5	&$<-$12.5		&5,9\\	% CLINES:	uv CI+II lines, koester+ 1982, opt Swan band in UVES
0811$+$250	&7920  	&0.54		&$-$5.4	&$<-$12.2		&4\\		% CLINES:	opt Swan bands, coutu+ 2019
0856$+$331	&9490  	&0.87		&$-$3.5	&$<-$11.8		&4\\		% CLINES:	opt Swan bands, CI lines, dufour+ 2005
0913$+$103	&8410  	&0.53		&$-$5.2	&$  -$11.6		&4\\		% CLINES:	opt Swan bands, coutu+ 2019
0935$-$371	&9380  	&0.78		&$-$4.2	&$<-$12.3	 	&6\\		% CLINES:	opt Swan bands, dufour+ 2005
1015$+$088	&7580	&0.53		&$-$6.0 	&$<-$12.5		&4\\		% CLINES:	opt Swan bands, coutu+ 2019
1142$-$645	&7950 	&0.58		&$-$5.5	&$<-$12.9		&4\\		% CLINES:	opt Swan bands, dufour+ 2005
1149$-$272	&6440  	&0.56		&$-$6.7	&$<-$12.8	 	&4\\		% CLINES:	opt Swan bands, giamichelle+ 2012
1708$-$147	&9280  	&0.54		&$-$3.9	&$  -$11.7		&7,9\\	% CLINES:	uv CI lines, koester+ 1995, opt Swan band in UVES
1831$+$197   	&7120  	&0.56		&$-$6.2	&$<-$12.8		&4\\		% CLINES:	opt Swan bands, dufour+ 2005
1837$-$619   	&8500  	&0.5:			&$-$5.0	&$<-$12.5		&2,9\\	% CLINES:	uv CI lines, wegner 1983, opt Swan band in UVES
1917$-$077	&10400 	&0.62		&$-$5.8	&$<-$12.8		&5,10\\	% CLINES:	uv CI lines, wegner 1981, no lines in UVES (except He I)
2059$+$316   	&9100  	&0.66		&$-$5.0	&$<-$12.5	 	&4\\		% CLINES:	opt Swan bands, dufour+ 2005
2140$+$207   	&7520  	&0.50		&$-$6.3 	&$<-$12.8		&4\\		% CLINES:	opt Swan bands, dufour+ 2005
2147$+$280  	&11000 	&0.5:			&$-$7.0	&$<-$12.0		&2,10\\	% CLINES:	uv CI lines (faint), koester+ 1995, no lines in UVES (except He I?)
2154$-$512	&7190  	&0.60		&$-$4.4 	&$<-$12.8		&5\\		% CLINES:	opt Swan bands, broad, CH band, giamichelle+ 2012, DQP = polarized, not DQp!
2311$-$068	&7350  	&0.55		&$-$6.1	&$<-$12.5		&4\\		% CLINES:	opt Swan bands, dufour+ 2005
2317$-$173   	&10800 	&0.5:			&$-$6.5	&$<-$12.3		&2,10\\	% CLINES:	uv CI lines, wegner 1983, no lines in UVES (except He I?) similar to 2147+280
		
\hline
\multicolumn{2}{l}{Other DQ(Z) stars:}\\

0208$-$510	&8180	&0.59		&$-$4.8	&$<-$11.8		&8\\		% CLINES: Swan bands, optical, farihi+ 2013
0736$+$053	&7590	&0.55		&$-$5.9	&$-  $11.8		&4\\		% CLINES: Swan bands, optical, provencal+ 2002

\hline

\end{tabular}
\end{center}

{\em Notes and references}: 
(1) \citet{blouin2019};
(2) \citet{weidemann1995};
(3) \citet{sayres2012};
(4) \citet{coutu2019};
(5) \citet{giammichele2012};
(6) \citet{dufour2005};
(7) \citet{subasavage2017}; 
(8) \citet{farihi2013a};
(9) C$_2$ band detected in the optical for the first time;
(10) carbon detected only in ultraviolet.

\end{table}

\subsection{Target selection and observations}

The observed DQ stars are listed in Table~\ref{targs}, and taken from the literature based on favorable brightness and position in the sky.  Science targets were selected to have $T_{\rm eff}\la11\,000$\,K, and thus are primarily classical DQ stars with C$_2$ Swan bands, where trace atmospheric carbon is attributed to the dredge up of core material via successful modeling in helium-rich white dwarfs \citep{pelletier1986,koester2020}.  These cooler DQ white dwarfs were selected because they represent the bulk of all objects in this spectral class, and form a modest but significant fraction of the local white dwarf population \citep{hollands2018}.  One known DQZ star (0913+103; \citealt{wegner1985,kleinman2013}) was included in the study in order to have at least one confident detection of Ca\,{\sc ii} H and K, and to search for other metal species and thereby better assess the rare, externally polluted DQ stars.

The overall science program was executed in service mode as priority B during ESO periods 95, 96, and 97, where data were ultimately received for 23 out of 30 possible targets.  Each white dwarf was observed with UVES \citep{dekker2000} on the VLT, using the 1\,arcsec slit and $2\times2$ binning on-chip, thereby achieving a nominal resolving power $R\approx40\,000$.  Spectroscopy was performed using a standard dichroic configuration, with central wavelengths of 3900\,\AA \ on the blue side, and 5640\,\AA \ on the red side.  Stars were observed using two identical integrations taken in sequence, where the individual exposure times varied between roughly 300\,s and 1800\,s (typically 900\,s) in order to achieve signal-to-noise (S/N) $>20$ at 4000\,\AA.

Raw science and calibration frames for all targets were retrieved from the ESO archive and reduced using the standard {\sc reflex} UVES pipeline. Spectra were flat-fielded, bias- and dark-subtracted, extracted, and wavelength-calibrated following parameter optimizations recommended in the  {\sc reflex} UVES documentation.  For the DA+DQ visual binary 0935$-$371, frames were reduced and calibrated using the standard processes, however for spectral extraction the slit length and offset parameters were adjusted to isolate each star. As observations were taken in pairs, individual spectra were combined via a weighted average to increase the S/N, and then normalized in each arm. For the entire sample, S/N values of $10-70$ in the blue, and $20-120$ in the red were found over the wavelength ranges 3500--3600\,\AA \ and 5200--5300\,\AA \ respectively. To provide confidence in the {\sc reflex} reductions, the ESO archive phase 3 spectra for all targets were retrieved and combined in an identical manner, where it was found that both the S/N and wavelength RMS values agree within 1 per cent.

%%% CA REGION FOR DQ(Z)/DZ %%%
\begin{figure}
\includegraphics[width=\columnwidth]{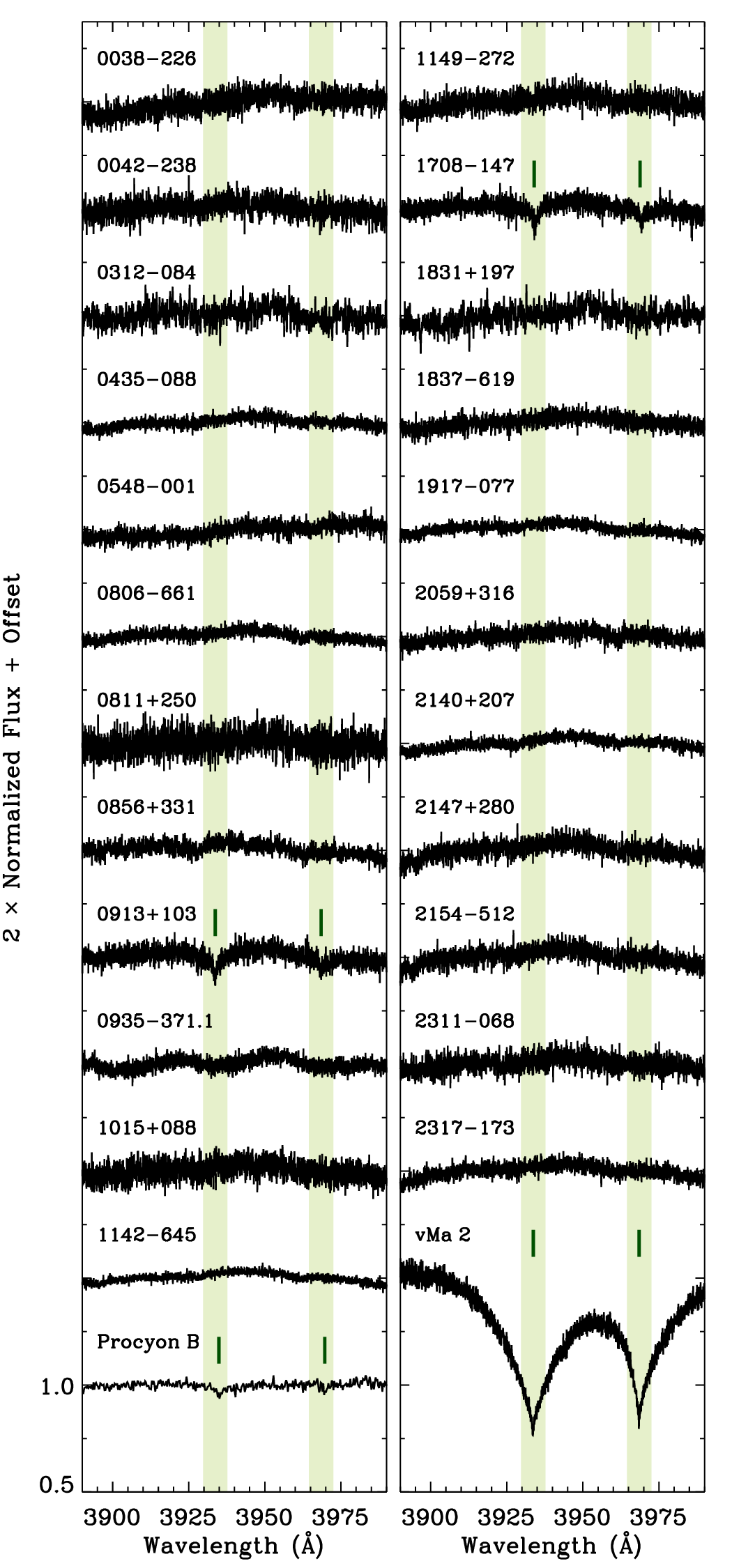}
\vskip 0pt
\caption{Spectra of all UVES targets in the region of Ca\,{\sc ii} H and K, highlighted by the shaded regions in each panel.  Spectra are presented at twice the normalized flux level, so that the vertical offset between spectra represents 50 per cent depth.  For all but two targets -- one previously known (0913$+$013) and one new discovery (1708$-$147) -- there are no indications of metal pollution via this most sensitive, ground-based diagnostic \citep{zuckerman2003}.  The spectra exhibit fringing, which is common in UVES data owing to a light path difference between the sky and internal flat lamp; otherwise, the spectra are featureless at these wavelengths.  Data for both 0312$-$084 and 1831$+$197 are smoothed by five binned pixels so that the noise is broadly similar to other targets.  At the bottom left is the DQZ star Procyon\,B, whose spectrum was obtained using {\em HST} STIS \citep{provencal2002} and reduced independently \citep{farihi2013a}.  At the bottom right is the prototype DZ star vMa\,2 as observed by UVES \citep{napiwotzki2003}, where the Ca\,{\sc ii} H and K features are dramatic, even at the relatively low abundance of [Ca/He] $= -10.0$ \citep{dufour2007}.
\label{uves}}
\end{figure}

%%% FIGURE MDOTS and HRD1 %%%
\begin{figure*}

\includegraphics[width=\columnwidth]{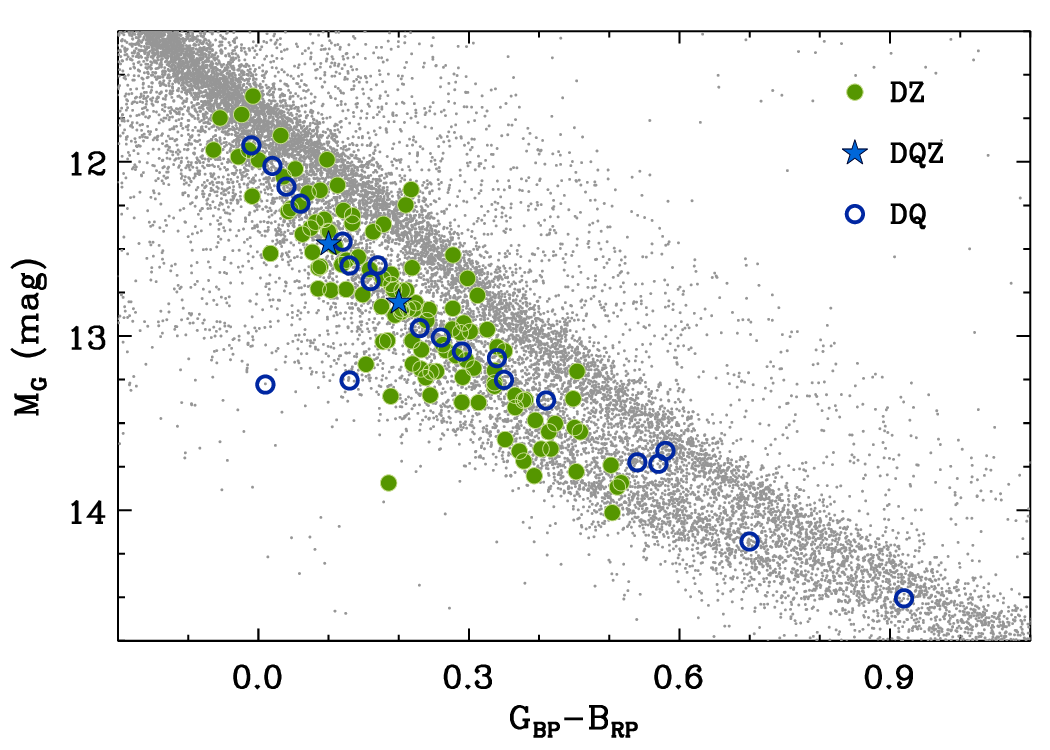}
\includegraphics[width=\columnwidth]{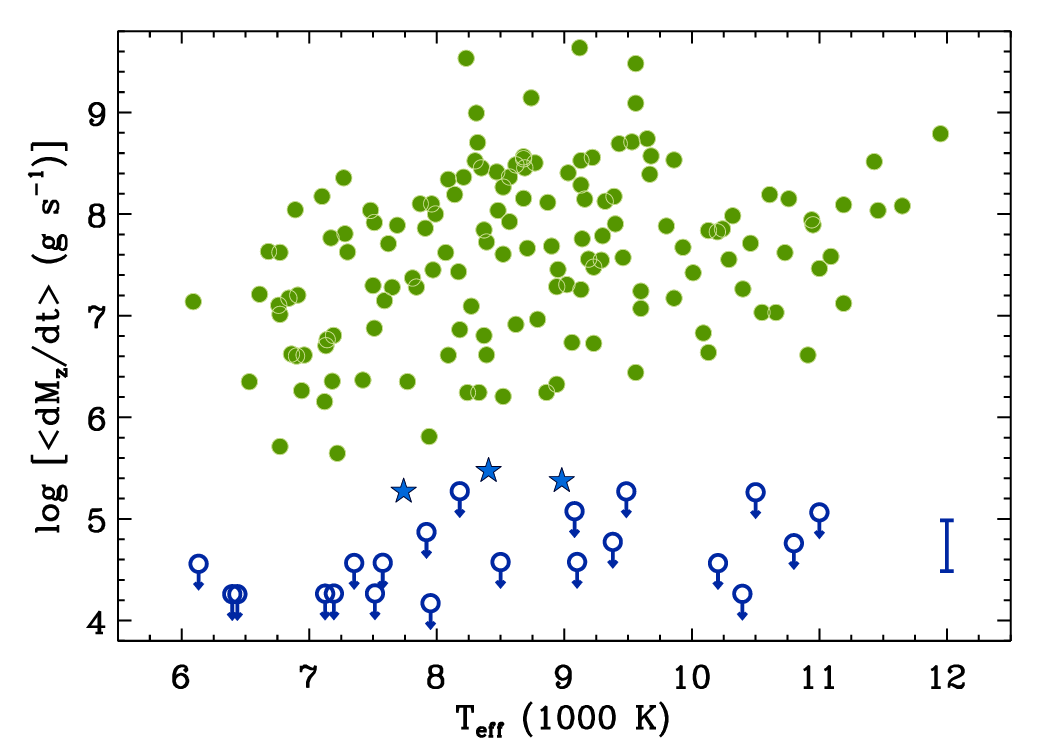}
\vskip 0pt
\caption{The similar yet disparate characteristics of the cool, helium atmosphere white dwarfs of spectral classes DQ and DZ.  {\em Left}:  The evolutionary tracks of the Hertzsprung-Russell diagram, based on {\em Gaia} EDR3 recommendations \citep{gaia2021}, where roughly 23\,200 stars are plotted as grey points.  The UVES sample of DQ target stars are plotted as open blue circles where no Ca is detected, and as filled blue stars for Ca detections (DQZ).  Filled green circles show DZ white dwarfs \citep{dufour2007}, where only those stars with a $20\upsigma$ or higher detection of their parallax are plotted.  {\em Right}:  Time-averaged accretion rates and upper limits for the DQ(Z) stars from Table~\ref{targs}, plotted alongside the same inferences for DZ stars from the SDSS \citep{farihi2010b}.  {\em The population of DZ stars and DQ(Z) stars exhibit no overlap in metal accretion rates}, where $\langle \dot M_{\rm DZ} \rangle /  \langle \dot M_{\rm DQZ} \rangle \approx 800$, and $\langle \dot M_{\rm DZ} \rangle /  \langle \dot M_{\rm DQ} \rangle > 3000$.  While the least metal-rich DZ stars shown here may be limited by the spectroscopic sensitivity of the SDSS, the most metal-polluted DQ stars are nothing like their DZ counterparts.  Note that two of the plotted stars were known to be DQZ prior to the inception of this survey; Procyon\,B \citep{provencal2002} and 0913+103 \citep{kleinman2013}.  The error bar in the lower right corresponds to a typical, $\pm0.25$\,dex uncertainty in [Ca/He] for DZ stars \citep{coutu2019}.
\label{mdots}}
\end{figure*}

\subsection{Abundances and inferred accretion rates}

Abundances\footnote{Abundance notation in this paper corresponds to [Z/He] = $\log(n_{\rm Z}/n_{\rm He})$.} and upper limits for Ca\,{\sc ii} were derived using state-of-the-art models of helium atmosphere white dwarfs, and specifically for stars that are enhanced in either carbon or metals typically found as external pollutants \citep{coutu2019}.  This was achieved by the method of line profile fitting, where, after a good fit was approximated, models were recalculated with $\pm0.3$\,dex abundances to determine the best fit, or upper limit (by inspection).  Basic stellar parameters such as mass, effective temperature, and carbon abundance were taken from the literature and are listed in Table~\ref{targs}, together with the results for calcium.

For each DQ star in the UVES sample, the [Ca/He] abundance determination or upper limit was converted into a time-averaged accretion rate following standard methodology.  For helium atmosphere white dwarfs with temperatures in the range considered here, metal sinking timescales are sufficiently long that a steady state cannot be inferred with any confidence \citep{koester2009a}.  However, the steady-state accretion rate is equivalent to a time-averaged value over a single sinking timescale, and is sufficiently informative in several contexts \citep{farihi2009a,farihi2016a,hollands2018}.  This is simply the mass of a particular metal in the outer, fully-mixed layers of the star, divided by the diffusion timescale for this element \citep{dupuis1993}.  To estimate the total mass accreted based on a single element, a correction factor is applied; in this case Ca is assumed to be 0.016 of the total mass as in the bulk Earth \citep{allegre1995,farihi2012b}.  Up-to-date diffusion calculations for pure helium atmospheres with no convective overshoot have been used \citep{koester2020}.

\section{Pollution inferences and implications}

The following section makes use of the metal abundances and upper limits from the UVES sample, as well as from a larger samples of DQ stars available from the Sloan Digital Sky Survey (SDSS).  While the UVES observations represent the most sensitive search for Ca\,{\sc ii} in DQ stars performed to date, the sample size is only 23 objects.  In contrast, there are an order of magnitude more DQ white dwarfs with SDSS spectroscopy, and although these data are somewhat less sensitive to Ca\,{\sc ii} (Section~3.3), the resulting detection frequency is more robust than for the smaller UVES sample.  First, accretion rates and upper limits are inferred for the white dwarfs with UVES spectroscopy (and for two DQ stars observed with {\em HST}).  Second, pollution frequencies are discussed in the context of robust samples with available SDSS spectroscopy.  Third, the combined evidence is evaluated and compared with what is known for the wider population of polluted white dwarfs.

\subsection{Accretion rates: UVES}

Figure~\ref{mdots} explores the inferred total (time-averaged) accretion rates for the observed DQ stars, alongside a sample of DZ stars with similar effective temperatures 
\citep{dufour2007,farihi2010b}.  Both sets of stars are understood to have helium-dominated atmospheres, and can be seen to occupy the corresponding evolutionary track on the Hertzsprung-Russell diagram shown in the Figure.  Although there appear to be a few DQ stars that are outliers in $G_{\rm BP}-G_{\rm RP}$ color, this is simply the result of objects with strong Swan band absorption in the blue, and not from any actual difference in the range of temperatures or cooling ages.  The two DQZ stars in the UVES sample lie directly on the helium atmosphere cooling sequence and do not otherwise distinguish themselves.

In stark contrast to their shared effective temperatures and luminosities, in terms of metal accretion rates, {\em the DQ and DZ stars show no overlap whatsoever}.  It may appear that there are no DZ stars with abundances and inferred accretion rates as low as the DQ(Z) objects, but there has not yet been any search for weak metal lines in DC white dwarfs using large telescopes and high-resolution spectroscopy.  The conventional detection of DZ stars is straightforward using low-resolution spectroscopy of modest quality, based on the fact that cool helium atmospheres are largely transparent owing to a lack of major opacity sources \citep{koester2009a}.  Notably, the Ca\,{\sc ii} lines in vMa\,2 were detected with photographic plate technology over 100 years ago \citep{vanmaanen1917}, and, as can be seen in Figure~\ref{uves}, using an 8\,m class telescope and a high-resolution spectrograph is clearly overkill, even for [Ca/He] $=-10.0$ ($\dot M_{\rm z} = 1\times10^7$\,g\,s$^{-1}$).  For context, DZ stars with metal abundances 30 times smaller than vMa\,2 are known, and the corresponding Ca\,{\sc ii} line strengths are readily detected using the $R\approx2000$ resolving power of SDSS spectroscopy \citep{dufour2007,coutu2019}.  

The use of the VLT and UVES, together with a sample size of 23 targets, demonstrates conclusively that the pollution in DQZ stars never approaches the levels typically attained in DAZ, DBZ, and DZ stars.  For systems that are likely in accretion-diffusion equilibrium (i.e.\ a steady state), instantaneous accretion rates can exceed $10^9$\,g\,s$^{-1}$ \citep{farihi2016b}, and roughly speaking, that is 5600 times higher than the rates inferred for polluted DQ white dwarfs.  And while steady-state accretion cannot be assumed for helium-rich atmospheres, if the time-averaged mass accretion rates are compared between the DQZ and the DZ stars, the top accretors in each class are separated by a factor of 18\,000(!).  Furthermore, most DQ stars have only upper limit metal abundances and accretion rates, and therefore the gulf between these two populations is vast.

\subsection{Pollution frequency: SDSS}

Although Ca\,{\sc ii} in the stars 0913$+$103 and 1707$-$147 was detected with UVES, these lines should be detectable in DQ stars via SDSS spectroscopy, for the same reasons outlined above for DZ stars; transparent atmospheres and broad lines.  Because DZ and DQ stars both have atmospheres dominated by helium, and similar effective temperatures, the much larger SDSS datasets of DQ white dwarfs will result in lower yet comparable sensitivity to Ca\,{\sc ii} absorption.  For this purpose, two SDSS spectroscopic samples were analyzed, where DQZ candidates were visually inspected, all of which are shown in Figure~\ref{sdss}.

First, there are 164 stars designated as DQ (160) or DQZ (4) in the SDSS DR7 white dwarf catalog, ignoring those spectral types deemed uncertain \citep{kleinman2013}.  On further scrutiny, three of these four are genuine DQZ (J091602.83$+$101109.7, J133205.62 +274003.9, J153447.54$+$414559.4), and one is likely a false positive with noisy data (J140256.39+111332.3; Figure~\ref{sdss}).  This represents 3:164 or $1.8${\raisebox{0.5ex}{\tiny$^{+1.7}_{-0.6}$} per cent of DQ stars that are DQZ.  Second, a sample of 221 DQ white dwarfs with reliable SDSS spectroscopic identifications based on S/N, and robust {\em Gaia} parallaxes were taken from the literature \citep{koester2019}, where each spectrum was examined for the possible presence of Ca\,{\sc ii} lines.  Of these, there are four DQZ stars; two are in common with the DR7 catalog, plus the objects J075230.82$+$444749.9 and J131953.49$+$084422.0, yielding 4:221 or $1.8${\raisebox{0.5ex}{\tiny$^{+1.4}_{-0.5}$} per cent.  During these visual searches, weaker candidates with noisy spectra were also considered, and the strongest of these are shown in Figure~\ref{sdss}; they are not considered further, and even if real the preceding frequency statistics would not be significantly altered.

There is a modestly larger sample of DQ stars that has been analyzed by \citet{coutu2019}.  The catalog is comprised of all stars designated as DQ and DZ in the Montr\'eal white dwarf database and thus prone to selection effects, but is largely comprised (80 per cent) of sources with SDSS spectroscopy.  Despite any possible biases within this heterogeneous sample, the numbers are revealing.  Of the 319 DQ white dwarfs studied, only four DQZ stars are noted; Procyon\,B and three SDSS objects re-evaluated in Figure~\ref{sdss}, where J090051.91$+$033149.3 is rejected here due to insufficient signal.  The remaining three sources with metal lines represent 3:319 or $0.9${\raisebox{0.5ex}{\tiny$^{+0.9}_{-0.3}$} per cent of DQ stars.  It is also noteworthy that in the same database, there are 1023 DZ white dwarfs with no evidence for carbon enrichment \citep{coutu2019}.  Again keeping in mind that this catalog has selection effects, it seems that external pollution favors helium atmospheres without carbon at a ratio of 1026:3 or in 99.7 per cent of cases.

%%% CA REGION FOR SDSS DQ(Z)/DZ %%%
\begin{figure}
\includegraphics[width=\columnwidth]{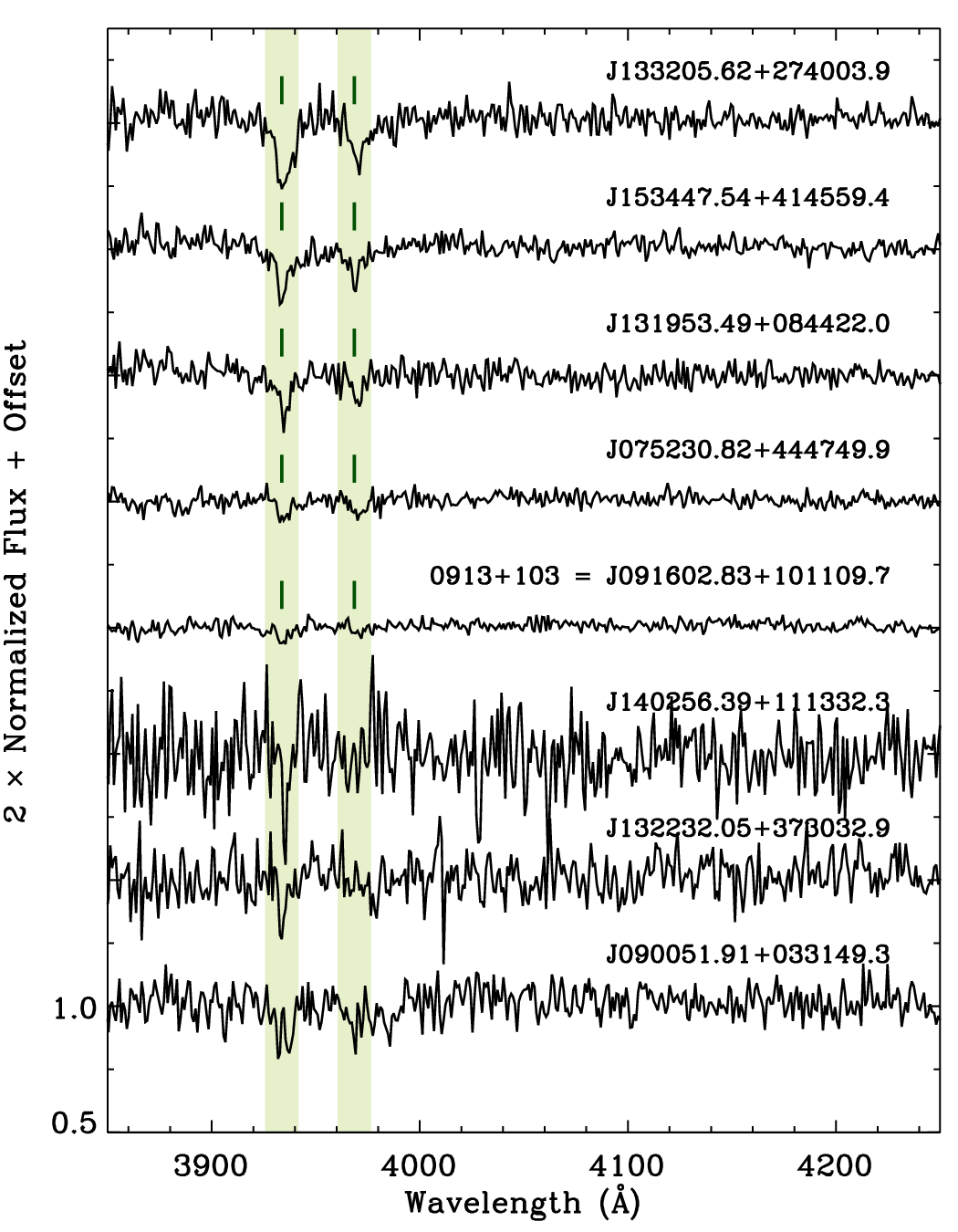}
\vskip 0pt
\caption{An approximate sequence of stronger to weaker Ca\,{\sc ii} absorption lines in DQ white dwarfs with SDSS spectroscopy.  All spectral data are displayed analogously to Figure~\ref{uves}, at twice the normalized flux level, so that the vertical offset between each spectrum represents 50 per cent depth.  From top to bottom, the first five spectra are considered to have confident detections of both the Ca\,{\sc ii} H and K features, where these are marked with solid vertical lines.  One of these was also observed with UVES; 0913+103 = J091602.83 +101109.7 \citep{kleinman2013}.  The final three spectra towards the bottom are considered too noisy, and these are likely false positives.  Altogether, these sources represent around 2 per cent of DQ stars with comparable spectroscopy. 
\label{sdss}}
\end{figure}

%%% 0913+103 COMPARISON%%%
\begin{figure}
\includegraphics[width=\columnwidth]{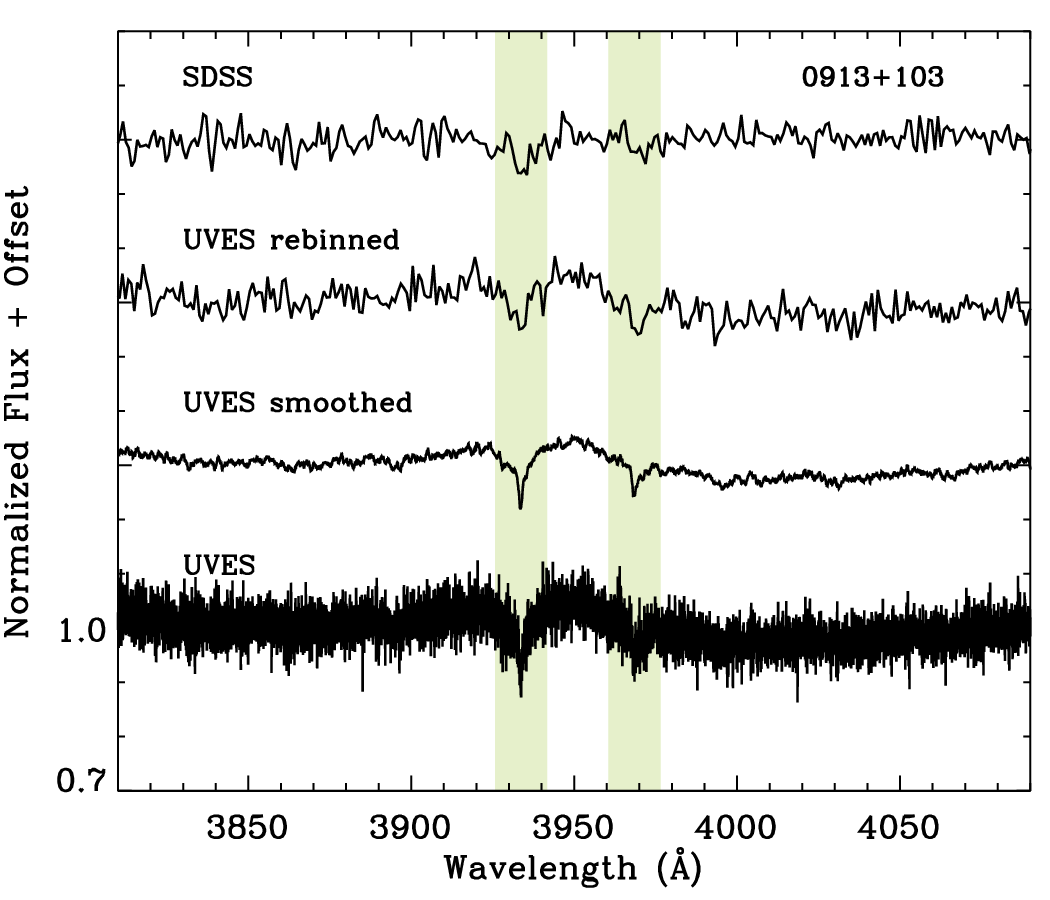}
\vskip 0pt
\caption{A side-by-side comparison of the Ca\,{\sc ii} H \& K lines detected in 0913+103 using UVES and SDSS spectroscopy, where all spectral data are normalized and offset by a constant of 30 per cent depth.  The resolving power of the SDSS (BOSS) spectrograph is $R=1650$ at these wavelengths, and the spectrum is shown at the top.  The UVES spectrum has resolving power $R=41\,400$, and is shown without modification at the bottom.  In between these are shown two modified versions of the UVES data; a rebinned spectrum that has the same number of wavelength points as the SDSS spectrum, and a smoothed spectrum that averages over a width equal to the rebinning factor.  The rebinned UVES and SDSS data are comparable, but the intrinsically weaker, Ca\,{\sc ii} H line is better detected in UVES.
\label{0913}}
\end{figure}

\subsection{0913+103: a bridge connecting UVES and SDSS}

The previous section takes advantage of large samples of DQ white dwarfs for statistical purposes, but a priori, $R\approx2000$ spectroscopy with the 2.5\,m SDSS telescope may not be readily comparable to $R\approx40\,000$ data from the VLT.  However, and perhaps surprisingly, the UVES target 0913+103 is weakly but clearly detected as a DQZ in the SDSS with designation J091602.83$+$101109.7, and was selected for the VLT survey based on this foreknowledge (Section~2).

Figure~\ref{0913} compares the Ca\,{\sc ii} lines detected in both SDSS and UVES, where the broad width of the weak lines -- K in particular -- enables their detection at moderately low spectral resolution.  The SDSS spectrum has an average S/N $=53$ (per resolution element) over its entire wavelength range from an exposure that is close to 3600\,s (4 integrations).  Although the UVES exposure time is comparable at 3960\,s (2 integrations), the raw spectrum has only S/N $=23$ in the blue arm where Ca\,{\sc ii} is found.  In the Figure, the UVES data are rebinned to the same number of points contained in the SDSS wavelength array, yielding a spectrum that is remarkably comparable in terms of S/N.  However, in these rebinned data, the H line appears more robustly detected, and both the H \& K features have modestly greater line depths, a likely result of the original higher resolution.

This example demonstrates that, owing to the intrinsic width of the metal absorption features in cool, helium atmosphere white dwarfs -- exemplified by vMa\,2 in Figure~\ref{uves} -- the detection of DQZ stars using SDSS spectroscopy is broadly comparable to their detection with UVES, at least to the same degree as found in 0913+103.  Using this DQZ white dwarf as a benchmark, it can be inferred that abundances and time-averaged accretion rates as low as [Ca/He] $=-11.6$ and $\dot M_{\rm z} = 3\times10^5$\,g\,s$^{-1}$ are likely detectable, and thus absent in the $N>200$ DQ stars in the SDSS.  This conclusion adds considerable weight to the stark difference in pollution between the DQ and DZ white dwarfs.

\subsection{DQZ stars vs. other polluted white dwarfs}

In Table~\ref{rates}, the different types of polluted white dwarfs are compared by 1) $f_{\rm z} \equiv$ the frequency at which pollution is detected and 2) $\langle \dot M_{\rm z} \rangle \equiv$ a typical mass accretion rate inferred for a given spectral class.

\begin{enumerate}

\item{For the spectroscopic class of DA stars (= hydrogen Balmer lines strongest, typically but not always hydrogen-dominated atmospheres), pollution frequency and accretion rate statistics are taken from \citet{zuckerman2003} and \citet{koester2014}.  Polluted spectral types include DAZ and DZA, where the latter implies the metal lines are the stronger than the Balmer lines.}

\item{For the spectroscopic class of DB stars (= He\,{\sc i} lines strongest, helium-dominated atmospheres), the occurrence rate of pollution and accretion statistics are taken from \citet{zuckerman2010}. Polluted spectral types include DBZ and DBAZ, where the latter indicates weak hydrogen lines, and even weaker metal lines.}

\item{For the spectroscopic class of DC stars (=  continuum only, no lines, typically but not always helium-dominated atmospheres), pollution statistics and accretion rate inferences are taken from \citet{farihi2010b} and \citet{hollands2018}.  Polluted spectral type is DZ.}

\end{enumerate}

 %%%TABLE RATES%%%
\begin{table}
\begin{center}
\caption{White dwarf pollution as a function of spectral class.\label{rates}}
\begin{tabular}{@{}r@{\hspace{2pt}}lrlc@{}}

\hline

\multicolumn{2}{c}{Spectral Class}	&$f_{\rm z}$	&$\langle \dot M_{\rm z} \rangle$	&Ref\\
&							&			&(g\,s$^{-1}$)					&\\

\hline

DAZ /	&[ DA + DAZ ]	&0.25		&$\sim10^8$					&1,2\\
DBZ / 	&[ DB + DBZ ]	&0.30		&$\sim10^8$					&3\\
DZ /		&[ DC + DZ ]	&0.28		&$\sim10^8$					&4,5\\

DQZ / 	&[ DQ + DQZ ]	&$<0.09$		&$\lesssim10^{5}$				&6\\
DQZ / 	&[ DQ + DQZ ]	&$0.02$		&$\lesssim10^{5.5}$				&7\\

\hline

\end{tabular}
\end{center}

{\em References}: 
(1) \citealt{zuckerman2003}; 
(2) \citealt{koester2014}; 
(3) \citealt{zuckerman2010}; 
(4) \citealt{farihi2010b}; 
(5) \citealt{hollands2018};
(6) This study, based on UVES;
(7) This study, based on SDSS.

\end{table}

The UVES sample studied here consists of 23 targets, including a known DQZ, and one new discovery.  The nominal fraction of externally polluted DQ stars is thus 1:22 or $4.5${\raisebox{0.5ex}{\tiny$^{+9.1}_{-1.4}$} per cent.  It can be argued, however, that the previously known DQZ is actually a part of the UVES sample, because it is possible it could have been a selected target without prior knowledge.  In that case, the pollution frequency for DQ stars could be as high as 9.1\, per cent.  However, on the basis of the much larger, unbiased SDSS samples of DQ white dwarfs discussed above, it is far more likely the true fraction is substantially smaller, and closer to 2 per cent.

As can be seen from Table~\ref{rates}, there is a general coherence in the picture of polluted white dwarfs among the spectral classes DA(Z), DB(Z), and DC/Z, where this is now a well-understood story of dynamically active planetary systems in the post-main sequence \citep{veras2016}.  Most of the aforementioned pollution frequencies and accretion rate inferences belong to studies more than a decade old, and in the intervening years those interpretations have been strengthened.  After cooling for 20\,Myr or so (during which the sensitivity to external pollution is poor, owing to hotter atmospheres; \citealt{koester2014,barstow2014}), a given white dwarf may exhibit metals at the observed frequency as a fraction of its remaining cooling lifetime (i.e.\ a duty cycle), or a subset of the entire population may stay constantly polluted, while others are never enriched by remnant planetary systems.

With a few notable exceptions, polluted white dwarfs are a population dominated by isolated stars \citep{wilson2019a}, together with a small fraction of wide binaries where each star has evolved in effective isolation \citep{zuckerman2014}.  Although stars with helium-rich atmospheres often have time-averaged accretion rates that exceed those inferred from DAZ stars in accretion-diffusion equilibrium \citep{girven2012,farihi2012b}, it is still the case that typical rates are comparable with those of their hydrogen-rich counterparts.

This story breaks down completely for DQ stars; not only is pollution rare, but when it is present it is stunted by several orders of magnitude.  The frequency of external pollution in DQ stars is drastically lower than for known polluted white dwarfs in Table~\ref{rates}.  On top of this stark difference in the frequency of detected metals, the discrepancy between the mass accretion rates is also severe; in fact, there is no overlap between these populations.  Unless DQ stars transform into spectral type DZ during external pollution events, it is highly improbable that they are directly related to, or descended from DA or DB white dwarfs that frequently retain active planetary systems.

\section{Additional anomalies}

In the previous sections, it was shown that DQ stars do not share any characteristics in common with either hydrogen or helium atmosphere stars that commonly evidence their remnant planetary systems via external pollution.  However, the striking lack of metal pollution in DQ stars is just one empirical perspective that suggests a distinct origin for these white dwarfs.  In this section, several other of their likely or potentially disparate characteristics are explored in detail.

\subsection{Lack of core dredge-up in DZ white dwarfs?}

While the statistics have been clear for decades, the present study reiterates the fact that DZ and DQ stars are essentially two mutually exclusive populations, and determines the size of the intervening gulf in terms of atmospheric metals.  But, importantly, there are two independent aspects that underly this lack of common spectral features; the fact that DQ stars are rarely DQZ (Section~3), and separately that DZ stars essentially never manifest as DZQ (\citealt{bergeron1997}; see below).  Assuming both are the result of single star evolution, then why don't any helium-rich white dwarfs with planetary system pollution, also dredge up their interior carbon?  In a scenario of isolated stellar evolution, the unavoidable interpretation is that these two spectral classes have distinct interior structures.  That is, there is no plausible mechanism for a planetary system orbiting a DZ white dwarf to prevent the host star from dredging up carbon from its core.

That being said, for a range metal abundances in DZ white dwarfs it has been demonstrated that atmospheric carbon bands, that might otherwise result from dredge-up, can be suppressed or masked \citep{blouin2022}.  Updated modeling shows that the C$_2$ Swan band opacity decreases significantly with increasing metal abundance associated with external pollution, and thus, in principle, a star that appears spectrally of type DQ, may in principle transform into spectral type DZ.  While certainly plausible for a range of parameters, roughly one quarter of the modeled DZQ spectra appear detectable with sufficient S/N.  Such an empirical test should discover if this potential mechanism actually operates.

It is well-established that core material dredge-up is most efficient when the outer envelope of helium is somewhat reduced compared to the predictions of standard stellar evolution.  The earliest calculations of carbon dredge-up indicated that efficiency peaked with helium mass fractions $-4.0 \lesssim \log q({\rm He}) \lesssim -3.5$ \citep{pelletier1986}, while a decade later, carbon enhancements were produced with $-3 \lesssim \log q({\rm He}) \lesssim -2$ \citep{macdonald1998}.  More recently, a similar range of helium layer masses has been corroborated, and with predictions that mostly bracket the classical, observed [C/He] sequence as a function of effective temperature \citep{dufour2005}, where state-of-the-art models indicate that $-3.5 \lesssim \log q({\rm He}) \lesssim -2.5$ best reproduces this abundance pattern \citep{koester2020}.  However, there are two known [C/He] sequences at the cool end \citep{dufour2005,koester2006b}, one of which requires even smaller helium layers $\log q({\rm He}) \lesssim -3$ \citep{coutu2019}, and the warmer DQ stars require drastically lower $\log q({\rm He}) \lesssim -5$ \citep{koester2020}.

It is well known that some DQ white dwarfs exhibit carbon features in the optical, while others do so only in the ultraviolet, and that this can be ascribed to differences in helium mass fractions rather than effective temperature \citep{weidemann1995,blouin2023b}.  In fact, as noted in Table~\ref{targs}, several of the DQ targets were found to have an optical C$_2$ feature for the first time in their UVES spectra (0806--661, 1708--147, 1837--619), while others only exhibit carbon features in the ultraviolet (1917--077, 2147+280, 2317--173) despite the new, highly-sensitive observations.  Therefore, it is possible or even likely that there is a continuous distribution of helium layer masses \citep{koester2020}, where thinner layers will lead to optical carbon features, modestly thicker layers result in ultraviolet carbon features only, and if sufficiently thick, no carbon will be effectively dredged.

It should be mentioned that there is one DZ star with ultraviolet carbon features (2216--657; \citealt{weidemann1989,wolff2002}), but the abundance has been shown to be consistent with carbon-rich planetary material, externally accreted, and thus not from core dredge-up \citep{swan2019b}.  And while it has been shown that the most extreme cases of metal pollution can easily mask carbon features in both the optical and ultraviolet, in general this may not be the case \citep{hollands2022,blouin2022}.  Therefore, the number of known cases where carbon features can be confidently ascribed to core dredge-up in DZ stars is likely (currently) zero.  That is, DZ stars may have helium mass fractions that are significantly and consistently larger than the range found for DQ white dwarfs.

There are two related points to be made from these successful and independent modeling efforts of [C/He] in DQ stars.  The helium atmosphere, DZ and DC stars appear to have larger $q({\rm He})$ so that core dredge-up is inefficient, and this stellar evolutionary outcome represents the majority of all cool, helium-rich white dwarfs (40:54 = 0.74 in the 20\,pc sample; \citealt{hollands2018}).  The modest to drastic reduction in $q({\rm He})$, necessary so that core carbon is effectively dredged in DQ stars, is sufficiently common that an explanation is needed, where possibilities are discussed in Section~5.

\subsection{Lack of circumstellar material orbiting DQ stars}

To date, all white dwarfs suspected or confirmed to have orbiting circumstellar disks associated with planetary debris are also observed to have atmospheres polluted by heavy elements \citep{farihi2016a}.  There are currently eight systems exhibiting irregular transits from their debris disks, all of which are polluted with the exception of a few, where available data are insufficiently sensitive \citep{vanderbosch2020,guidry2021}.  Confirmation of photospheric metals may require a high spectral dispersion sufficient to resolve or rule out interstellar absorption components \citep{koester2005a,zuckerman2013}, or ultraviolet spectroscopy for modest to low metal abundances in hotter stars \citep{farihi2013b,xu2015}.

While {\em Spitzer} observations were ground-breaking for debris disk detections towards white dwarfs, over its 16 year lifetime, targets were strongly biased towards stars already known to be polluted.  Owing to this bias, DQ stars were not searched for infrared excess with comparable sensitivity, with one possible exception \citep{farihi2010a}.  However, unbiased studies using the {\em WISE} satellite have been able to probe much larger, unbiased samples \citep{debes2011,hoard2013,dennihy2017}, now including nearly every sufficiently bright white dwarf identified by {\em Gaia} and utterly blind to spectral type \citep{xu2020,lai2021}.  Optical spectroscopic follow-up of debris disk candidates is still ongoing, but so far there is not a single DQ star candidate (E.~Dennihy, private communication) among $N\gtrsim100$ white dwarfs where orbiting debris is detected either by infrared excess, optical emission or absorption lines, or transits.
 
The dearth of detected circumstellar material orbiting DQ white dwarfs may appear -- superficially -- redundant with the lack of atmospheric pollution, as there is a strong connection established between these two hallmarks of evolved planetary systems.  As a rule for white dwarfs, photospheric heavy elements require ongoing or recent accretion; that is, pollution implies closely-orbiting circumstellar material, where it is understood that the bulk of debris disks are simply undetectable owing to sensitivity limits \citep{rocchetto2015,bonsor2017}.  But the converse is not necessarily true, and planetary matter can exist in a wide range of orbits and sizes that would not necessarily lead to the chemical enrichment of the host star.  Little is known about the architecture of planetary systems orbiting polluted white dwarfs outside of a few tens of R$_{\odot}$ \citep{xu2013,farihi2014}, but there are at least two examples of the aforementioned scenario: there is a giant planet candidate, transiting the non-polluted white dwarf WD\,1856+534 (spectral type DC; \citealt{vanderburg2020}), and a 2500\,AU, planetary-mass companion orbiting WD\,0806-661 (spectral type DQ, Table~\ref{targs}; \citealt{luhman2012}).

As mentioned above, existing observations may be insufficiently sensitive to photospheric metals, yet able to detect circumstellar material.  For example, PG\,0010+280 is a $T_{\rm eff}\approx27\,000$\,K DA white dwarf with an apparent infrared excess from a debris disk, but lacks photospheric metal lines at optical wavelengths, presumably due to its relatively high effective temperature and opacity \citep{xu2015,walters2022}.  However, current models do not predict significant opacity in DQ stars, which are understood to have helium-dominated atmospheres, possibly with trace abundances of hydrogen that modestly impact models at the lowest effective temperatures \citep{koester2020}.  Atmospheric modeling of DQ white dwarfs already includes opacities for carbon molecules and atoms \citep{dufour2005,coutu2019}, and thus there is currently no other suspected opacity source that might otherwise mask metal lines if such heavy elements were indeed present in their atmospheres. 

Another means of masking the detectable signature of external pollution, where the detection of circumstellar matter might be more straightforward, would be to dilute the accreted material into a much larger stellar reservoir.  While speculative, if the mass of the outer, fully mixed layers of DQ white dwarfs were orders of magnitude larger than expected from current theory, then the strength of any metal absorption features would be correspondingly weaker than predicted, and the derived abundances commensurately smaller.  All else being equal, the DZ and DQ accretion rates (e.g.\ Figure \ref{rates}, treating upper limits as detections for heuristic purposes) could be made to agree, if the mass of the convection zones in DQ white dwarfs are roughly $3000\times$ more massive.  While this possibility is almost certainly pathological, in this case DQ stars could accrete from circumstellar disks that might be detected, yet the resulting pollution would be drastically diminished.  

However, no circumstellar matter has been detected towards DQ white dwarfs, and this partly mitigates the possibly of masking any pollution as typically observed in a DZ star.  Statistics on infrared excesses indicate that robust detections are rare below 9000\,K or cooling ages significantly longer than 1\,Gyr \citep{bergfors2014,farihi2016a}, and thus the lack of infrared excess is not the strongest constraint on the lack of orbiting material, but it is suggestive.

%%% KK19 DQM FiGURE %%%
\begin{figure*}
\includegraphics[width=\columnwidth]{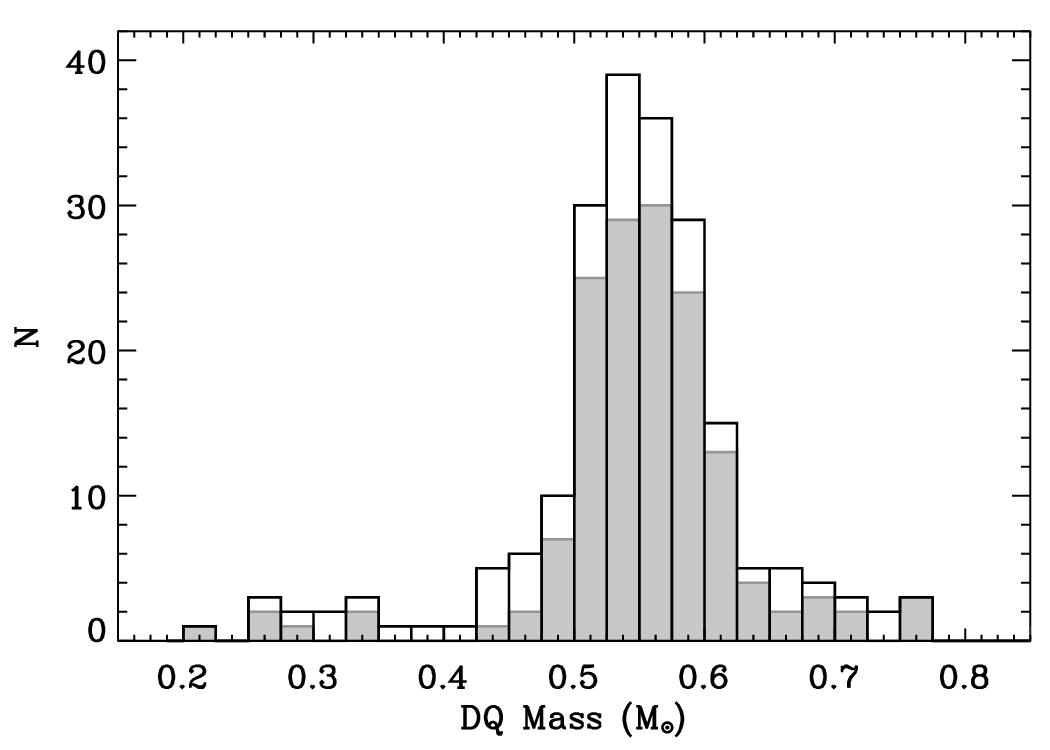}
\includegraphics[width=\columnwidth]{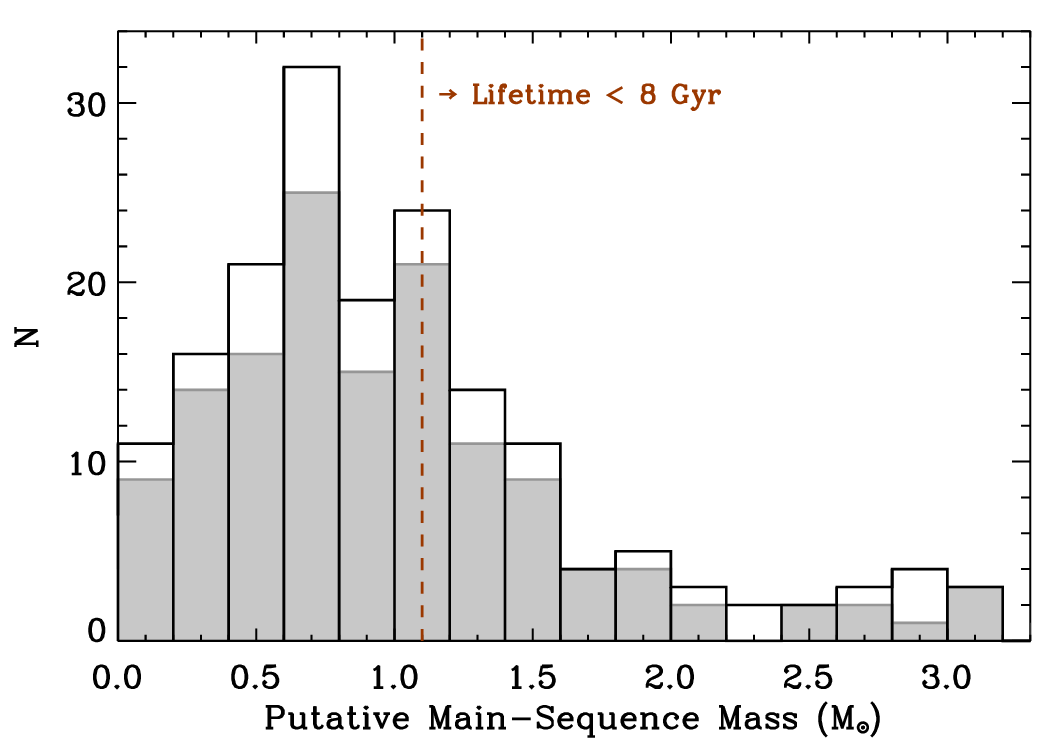}
\vskip 0pt
\caption{The left-hand panel plots the mass distribution within the 221 classical DQ white dwarfs with reliable SDSS spectroscopy and {\em Gaia} parallaxes \citep{koester2019}.   In the histogram, masses with fractional errors less than 10 per cent are plotted as unfilled bars (208 sources), and those less than 5 per cent are shown as filled bars (153 sources).  The low mass tail persists in those DQ white dwarfs with high quality data, where $M\lesssim0.45\,$M$_{\odot}$ is not possible for single star evolution within a Hubble time.  The right-hand panel implements a state-of-the-art, empirical initial-to-final mass relation \citep{cummings2018} to illustrate the likely distribution of main-sequence progenitor masses for these DQ white dwarfs, assuming single star evolution.  Most of these putative parent stars would have precursor lifetimes longer than 8\,Gyr ($M_{\rm MS}\lesssim1.1$\,M$_{\odot}$, based on MIST evolutionary models; \citealt{dotter2016}), which is inconsistent with expectations for thin disk stars \citep{bensby2014,kilic2017,sharma2019}.  The addition of a typical 1.2\,Gyr cooling age then amplifies the implied total ages of the DQ white dwarfs, suggesting they should be predominantly thick disk and halo stars if they evolved in isolation.
\label{mhist}}
\end{figure*}

\subsection{Observed lack of close DQ+dM binaries}

Low-mass main-sequence stars dominate the initial mass function, and the mass functions for companions to white dwarfs and to their A-type star progenitors \citep{farihi2005b,derosa2014}.  Thus, from a formation and evolutionary point of view, it is expected that white dwarfs with M dwarf companions are abundant, which is observed to be the case \citep{rebassa2007,hollands2018}.  Of these prevalent stellar pairs, based on both theory and observations, it is expected that roughly two-thirds will be in wide binaries where the original semimajor axis increased in response to post-main sequence mass loss, and around one-third will be in short-period orbits that resulted from common envelope evolution \citep{farihi2010c,nebot2011,ashley2019}.

Another property of DQ stars, that implies deviation from typical pathways of isolated stellar evolution, is the fact that there are no known examples in post-common envelope binaries with an unevolved companion (e.g.\ an M dwarf).  To characterize this deficit in more detail, the largest available spectroscopic catalog of white dwarf -- main sequence binaries was searched for any DQ white dwarfs.  This catalog is a compendium of all sources with composite (white dwarf + main-sequence star) spectra within the SDSS \citep{rebassa2012,rebassa2016}, and contains over 3200 entries, 91 per cent of which are spatially-unresolved, plus a small fraction of partly-resolved pairs that contribute their combined light into a single spectroscopic fiber.

The catalog was searched for all systems containing an M dwarf companion, which represents 98 per cent of all entries, and the primary spectral type was noted in each case.  Only confident spectral type identifications were used for both the white dwarf and the companion, and only those systems that are spatially unresolved.  Of the unresolved composites, 2263 have a white dwarf of spectral type DA (strongest lines from H), 93 contain a DB star (strongest lines from He\,{\sc i}), and 70 host a DC white dwarf (no spectral lines).  Not a single DQ white dwarf is present in the catalog, including entries with uncertain spectral type identifications. 

Because DC and DQ white dwarfs share similar stellar properties -- with the likely exception of the atmospheric helium mass fraction -- one would expect a comparable number of each in the SDSS catalog, adjusted for their relative occurrence rates as field stars.  Within the 20\,pc sample, there are 29 DC and 15 DQ white dwarfs \citep{hollands2018}, and if their frequency in unresolved (including short-period) binaries were similar, one would expect close to 35 DQ spectral types within the composite binary catalog.  Therefore, the lack of DQ stars in this catalog is striking.  While identifying a weak C$_2$ 5165\,\AA \ feature may be difficult in low S/N spectra, and in the presence of flux from an M dwarf companion, many DQ white dwarfs exhibit deep and strong Swan bands that are unmistakable \citep{dufour2005}.

It is important to keep in mind that the stellar wind from a low-mass, main-sequence star can donate hydrogen to a helium-atmosphere white dwarf via wind capture, if the orbit is sufficiently close, as can be the case in post-common envelope binaries.  In this way, a helium atmosphere white dwarf with a sufficiently thin convection zone (or fully mixed layer) can be transformed into one where hydrogen is more abundant.  This scenario is likely responsible for the dominance of DA spectral types in close, white dwarf -- main sequence pairs, including the extensive SDSS catalog, and is consistent with the lack of any known, post-common envelope systems containing a helium atmosphere white dwarf with a thin convection zone (e.g.\ DB white dwarfs; \citealt{vandenbesselaar2005,rebassa2010}).  

To demonstrate the plausibility of this atmospheric transformation, consider a $T_{\rm eff}\approx20\,000$\,K, pure helium atmosphere white dwarf, which will manifest as a DB spectral type, and where the fully mixed outer layer has mass $q\lesssim10^{-10}$\,M$_{\odot}$\citep{koester2009a,koester2020}.  If this white dwarf accretes solar composition wind from a companion at a modest rate of $\dot M = 10^{-14}-10^{-16}$\,M$_{\odot}$\,yr$^{-1}$ \citep{pyrzas2012,parsons2012,ribiero2013}, hydrogen will come to dominate by number within a fraction of a Myr.  With sufficient sensitivity, a hydrogen(-enriched) atmosphere white dwarf accreting stellar wind should appear as a DAZ spectral type, and there are several well-documented cases among post-common envelope binaries \citep{zuckerman2003,debes2006,tappert2011}.

However, for the aforementioned, pure helium atmosphere case, once the convection zone deepens due to further cooling, any superficial hydrogen layer may be easily overwhelmed.  By the time such a white dwarf has cooled to $T_{\rm eff}\approx10\,000$\,K, the size of the fully mixed layer should be $q\sim10^{-5}$\,M$_{\odot}$, where this applies both to DQ white dwarfs and their helium atmosphere DC star counterparts \citep{koester2009a,koester2020}.  With convection zones of this depth, it would take several Gyr or longer for the atmosphere to become dominated by hydrogen that is captured from a companion stellar wind, with a strong dependence on the binary separation (i.e.\ wind capture rate).  Therefore, it is expected that some DC, DZ, and DQ white dwarfs should be hosts to post-common envelope, main-sequence companions.  Indeed, there are a handful of such systems known to host DC stars \citep{nebot2011,parsons2013}, and at least one candidate DZ accreting metals in a stellar wind from an M dwarf \citep{fajardo2016}.  But there are none with a DQ white dwarf.

In contrast, there are many examples of DQ stars in wider binaries.  Within the 20\,pc sample alone, there are several: Procyon\,B (F5IV+DQZ), GJ\,86B (K0V+DQ), LHS\,290 (DQp+dM), 2154-512 (DQ+dM), 1633+572 (DQ+dM), and 1917-077 (DBQ+dM).  Thus, the lack of DQ white dwarfs in close or post-common envelope systems is likely a clue to their origin, and consistent with the destruction of close stellar companions during prior evolutionary phases.

\subsection{Apparently low DQ white dwarf masses}

The left-hand panel of Figure~\ref{mhist} plots the mass distribution for the sample of 221 classical DQ white dwarfs discussed in Section~3.2 \citep{koester2019}.  The masses of DQ stars have been studied recently by two independent teams, based on sources with reliable parallax data from {\em Gaia}, finding that their average mass is roughly 0.05\,M$_{\odot}$ lower than those found for DA and DB stars \citep{coutu2019,bedard2022}, and also lower than DZ and DC white dwarfs \citep{koester2019}.  There have been suggestions that there may be model imperfections that prevent better agreement between DQ stars and other white dwarfs, but to date, no other studies have considered enhanced mass loss that may result from binary evolution.

It should be noted that two iconic DQ white dwarfs in the 20\,pc sample have masses derived independently of models, and these values are also lower than the typical $\langle M \rangle=0.62$\,M$_{\odot}$ found for spectral type DA and DB white dwarfs \citep{genest2019}.  In fact, the dynamical mass of Procyon\,B (0.59\,M$_{\odot}$) is in modest tension with the overall age estimated for this well-studied binary system  \citep{liebert2013,bond2015}.  Adopting a comprehensive, recent, and empirical initial-to-final mass relation, the predicted mass of Procyon\,B should be closer to 0.66\,M$_{\odot}$ \citep{cummings2018}.  Similarly, for the planet-hosting binary GJ\,86, the DQ white dwarf has a mass of 0.54\,M$_{\odot}$ \citep{zeng2022}.  And while it is likely this system is relatively old (either thick disk or somewhat younger; \citealt{fuhrmann2014}), any recent initial-to-final mass relation will yield a progenitor that should not have evolved off the main sequence in a Hubble time.  Therefore, it is possible that the masses of DQ stars are actually low compared to other white dwarfs, and not a result of model shortcomings, but as a consequence of their evolutionary history.

Among the 221 DQ white dwarfs plotted in Figure~\ref{mhist}, there are 23 objects with derived masses $M<0.45$\,M$_{\odot}$, where 16 of these have $\upvarpi/\upsigma_\upvarpi > 20$, and thus excellent astrometric data.  It is often the case that (apparently) over-luminous white dwarfs are two unresolved stars with comparable brightness, and therefore {\em Gaia} DR3 was searched in a 5\,arcsec radius around the positions of these 23 candidate low-mass DQ stars; none were found to be resolved into two sources.  However, two were found to have renormalized unit weight error (RUWE) suggestive of a possible or likely non-single star:  SDSS\,J094058.34+384350.2 (RUWE $=1.37$) and SDSS\,J125245.44+194311.2 (RUWE $=2.33$).  Thus, the inferred low masses of these 23 DQ white dwarfs may be partly accounted for as binary light sources, but the bulk currently lack indications of duplicity.

%%% KK19 CHIST FiGURE %%%
\begin{figure*}
\includegraphics[width=\columnwidth]{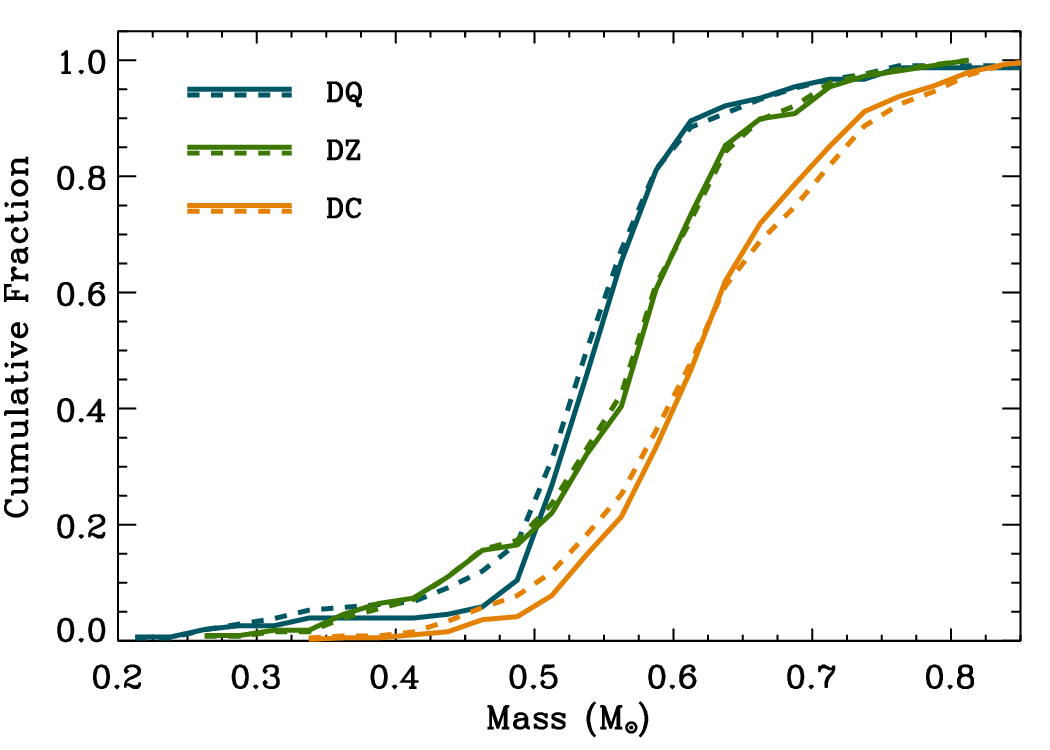}
\includegraphics[width=\columnwidth]{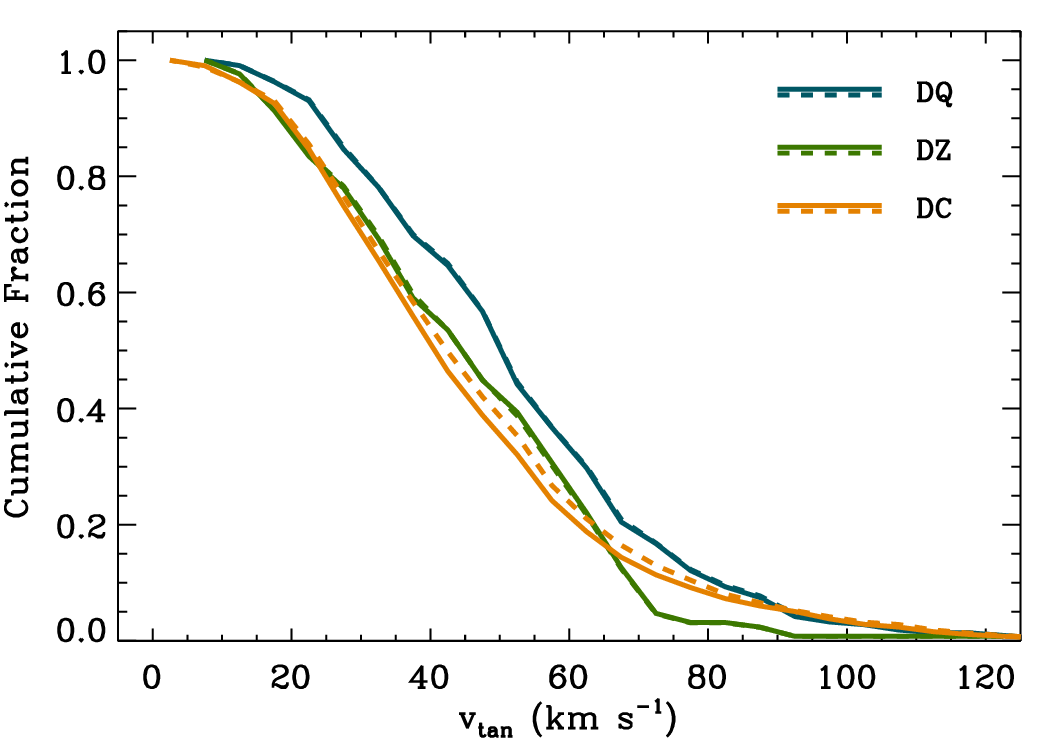}
\vskip 0pt
\caption{Cumulative mass and tangential speed histograms for the DQ, DZ, and DC stars with $T_{\rm eff}<10\,000$\,K analyzed in table 3 of \citet{koester2019}.  The dashed lines track the distribution for stars with better than $5\upsigma$ errors in mass or parallax (for $v_{\rm tan}$), while the solid lines follow the distributions for sources with errors better than $10\upsigma$.
\label{chist}}
\end{figure*}

Similarly, the argument is often made that a DQA-type spectrum must be two unresolved white dwarfs, one a DQ and one a DA star, but currently there is only verification of one such binary (GD\,73; \citealt{vennes2012}), where it is modestly overluminous relative to single star expectations \citep{gentile2019}, and has an RUWE consistent with a single astrometric source.  Of the 23 DQ white dwarfs with implied low masses, only one source shows unambiguous (yet weak) absorption from hydrogen in its spectrum (SDSS\,J150126.78+210056.9), but has an RUWE in the range considered normal for single stars, and no other {\em Gaia} sources nearby.  It can be argued that these low-mass DQ candidates are unresolved DQ+DC white dwarfs, but again one might expect a preponderance of high {\em Gaia} RUWE values, which is not yet observed.

Given that all DQ stars must have atmospheric carbon by definition, and thus carbon cores, it may seem oxymoronic to speculate on the possibility of low-mass members of this class.  Owing to the finite age of the Galaxy, white dwarfs with $M\lesssim0.45$\,M$_{\odot}$ cannot yet have formed by single star evolution.  Indeed the majority of low-mass white dwarfs are found to be single- or double-lined binaries \citep{marsh1995,brown2011a,rebassa2011}, whose evolution is thought to be terminated prior to helium ignition, and thus such stars would have helium cores as opposed to carbon cores.   

However, it is possible to form a low-mass carbon core in stars with $M_{\rm MS}\gtrsim2.3$\,M$_{\odot}$, where helium ignites non-degenerately \citep{bauer2021}.  Binary interactions would then be required to truncate stellar evolution and further core growth, in order to later be observed as a low-mass white dwarf with a carbon core.  While speculative, this is a distinct possibility for some DQ white dwarfs, as their binary properties are only weakly constrained at present, and their molecular bands are relatively insensitive to radial velocity changes induced by companions, especially compared to white dwarfs with atomic spectral lines.  The confirmation of any DQ white dwarf with $M<0.45$\,M$_{\odot}$ would be a strong indication of a binary origin.

The expectations for single star evolution are illustrated in the right-hand panel of Figure~\ref{mhist}, where main-sequence progenitor masses have been calculated using the previously mentioned initial-to-final mass relation.  Given the low masses of the DQ white dwarfs, it is thus not surprising that the predicted main-sequence predecessors are also of relatively low mass, but this important detail has so far escaped attention in previous studies \citep{coutu2019,koester2019,bedard2022}.  The age of thin disk stars should be less than around 8\,Gyr \citep{bensby2014,kilic2017,sharma2019}, but Figure~\ref{mhist} indicates that most single star parents of DQ white dwarfs would have main-sequence lifetimes longer than this.  For the DQ stars plotted in the Figure, $\langle T_{\rm eff} \rangle = 7800$\,K and $\langle M \rangle = 0.55$\,M$_{\odot}$, where white dwarf cooling would add an additional 1.2\,Gyr to their inferred total ages.  This picture would imply that DQ stars are dominated by thick disk and halo members, which, as shown below, is inconsistent with their kinematical and companion metallicity properties.

\subsection{Contrasts with DZ and DC stars}

It is useful to compare evolutionary indicators of DQ white dwarfs with their helium atmosphere counterparts, the DZ and DC stars.  While the mass distribution of DQ stars has been discussed above, and in particular the potential of a low-mass tail, this could simply be the result of unresolved binaries.  If that is indeed the case, and DQ stars are the product of isolated stellar evolution, then it is expected that other white dwarf classes would also exhibit the same basic properties in their mass distributions.  In Figure~\ref{chist}, the cumulative mass distributions are shown for the 221 DQ white dwarfs discussed above, together with SDSS spectroscopic DZ and DC samples from the same study \citep{koester2019}.  Only those DZ and DC stars with $T_{\rm eff} < 10\,000$\,K, and which thus overlap with the DQ white dwarfs, have been used in order to compare these three spectral classes of helium atmosphere stars using identical criteria.

The most obvious features of the cumulative mass histogram is that DQ white dwarfs have a larger fraction of lower masses, the DC stars have a notable fraction of higher masses, while the planetary system hosting DZ stars are somewhat in the middle of these two.  Another feature is the concentration of DQ white dwarf masses in the range $0.5-0.6$\,M$_{\odot}$, shown by a steepness in the distribution that is not observed for DZ or DC stars.  

One possible interpretation of this figure panel is as follows.  Planetary systems form and evolve commonly around the progenitors of DZ stars, which leave remnants with a given mass distribution as observed.  The DQ white dwarfs originate in a distinct population, with lower remnant masses (reasons explored below), and rarely formed planetary systems that lead to white dwarf pollution.  While speculative, the DC white dwarfs may descend from stars that are, on average, somewhat more massive than DZ star progenitors,  and which may indicate that planet formation is inhibited to some degree in higher-mass main-sequence stars.

The tangential velocity distributions tell a similar story but with less distinction between the DZ and DC white dwarfs.   The DZ stars appear to include few stars with relatively high velocities, consistent with planetary hosts coming from a primarily young disk population, where star formation likely benefitted from a comparatively metal-rich environment.  The DQ stars are modest outliers compared to their helium atmosphere, DZ and DC white dwarf counterparts, and seem to be missing stars at the lowest velocities, below around 20\,km\,s$^{-1}$.   This may imply the DQ white dwarfs are somewhat older than the DZ and DC stars.

Another possibility is that DQ white dwarfs have received significant velocity kicks from asymmetric mass loss, and which was not experienced by the DZ and DC stars.  Only modest white dwarf recoil, no larger than 1\,km\,s$^{-1}$, has been inferred using a large and pristine catalog of wide {\em Gaia} binaries, via comparison of those pairs with two main-sequence stars, versus those containing at least one white dwarf \citep{elbadry2018}.  However, based on the large separation between binary components in that study, this modest recoil may apply only to stars that evolved in effective isolation.  In contrast, close binary evolution may induce larger kicks that may account for the lack of low space velocities in the DQ stars.

The space motions of DQ white dwarfs are far from that expected for Population II stars, and neither are there any indications of a Galactic thick disk or halo origin.  Under the assumption of zero radial velocity, the spectral classes plotted in Figure~\ref{chist} have calculated mean space motions in the direction of Galactic rotation ($V$) given in Table~\ref{galvs}.  The three samples are without any kinematical biases, and were selected based on spectral classification via SDSS observations \citep{koester2019}.  As can be seen, the DQ stars may exhibit a modest fraction of older -- or kicked -- thin disk stars within their population, but for the most part, the DQ, DZ, and DC white dwarfs are members of the relatively young Galactic disk \citep{bensby2014}.  Thus, the low masses of the DQ white dwarfs, and the older ages implied by single star evolution, are not supported by their space velocities.

\subsection{Magnetism and rotation among DQ stars}

Magnetism has long been associated with DQ white dwarfs, including detections of strong fields \citep{schmidt1995,schmidt1999,vornanen2010}, as well as spectral indications of magnetism via modeling \citep{hall2008,blouin2019a}.  In a pioneering effort to characterize magnetism among all white dwarfs within 20\,pc \citep{bagnulo2021}, it has been recently found that DQ stars are likely outliers.  First, the nominal frequency of magnetism in DQ white dwarfs is higher than in other spectral classes.  Second, the actual frequency is likely to be higher, because the sensitivity is restricted to magnetic fields on the order of MG or larger, while spectropolarimetry can readily detect fields of 100\,kG or somewhat smaller in white dwarfs with atomic lines (i.e.\ DA, DB, or DZ stars).  Third, it is noteworthy that, in fact, the DQ stars have the highest magnetic fields of white dwarfs within 20\,pc \citep{bagnulo2021}.
  
Rapid rotation has been associated with some magnetic white dwarfs, especially in the scenario where close stellar binaries or mergers may generate the observed fields \citep{liebert2005,tout2008}.  Thus, relatively rapid rotation might be expected in some DQ white dwarfs, and appears to be well documented for the warmer examples of this spectral class \citep{dufour2008b,lawrie2013,williams2016}.  Based on these modest correlations, a more thorough investigation of the entire DQ spectroscopic class would likely identify additional magnetic examples, as well as stars rotating more rapidly than the tens to hundreds of hour spin periods typical for isolated white dwarfs \citep{hermes2017}.

%%%TABLE GALVs%%%
\begin{table}
\begin{center}
\caption{Space velocities in the direction of Galactic rotation.\label{galvs}}
\begin{tabular}{@{}ccc@{}}

\hline

Spectral		&$\langle V \rangle$	&$\upsigma_V$\\
Class		&(km\,s$^{-1}$)		&(km\,s$^{-1}$)\\

\hline

DQ			&$-13$			&30\\
DZ			&$-5$			&22\\
DC			&$-7$			&25\\

\hline

Thin disk		&$-15$			&20\\
Thick disk		&$-46$			&38\\

\hline

\end{tabular}
\end{center}

{\em Notes}:  Values for the thin and thick disk are taken from \citet{bensby2014}, and $v_{\rm rad}=0$ is assumed for all white dwarf calculations.

\end{table}

\section{Possible origins of DQ stars}

The previous section forms a list of independent yet circumstantial evidence that may help to constrain the origin of DQ white dwarfs.  And with the exception of the dramatic lack of metal pollution, the empirical data are suggestive but far from conclusive.  In this section, the various threads are combined and addressed with concrete hypotheses, which can be tested against the existing evidence, now and in the near future, and will hopefully set the stage to reveal the origin of all DQ white dwarfs; classical examples, warmer stars with atomic lines, spectral oddballs, as well as peculiar and magnetic examples.

\subsection{Summary of observations}

Below is a concise list of observational constraints, which can directly be compared to any hypothetical evolutionary possibilities.  Any successful hypothesis for the origin of DQ stars should be consistent with these data and indications, where Table~\ref{hypoths} tracks the various scenarios and their potential success in addressing each.

\begin{description}

\item{(1) DQ stars and DZ planetary hosts may be mutually exclusive populations, with hundreds of known examples of each. }\smallskip

\item{(2) There are no known DQ stars with infrared, optical, or transit indications of circumstellar, planetary material.}\smallskip

\item{(3) DQ stars are not found in post-common envelope binaries with unevolved companions.}\smallskip

\item{(4) The DQ spectral class is characterized by low $q({\rm He})$, where the bulk of members may have relatively low masses.}\smallskip

\item{(5) DQ white dwarfs are commonly magnetic, and likely more frequently or with stronger fields than other spectral classes.}\smallskip

\item{(6) The kinematics of DQ white dwarfs reveal they are only slightly older, or kicked, relative to DZ and DC stars.}

\end{description}

\subsection{Carbon is masked in metal-polluted stars}
It is plausible that external pollution onto DQ white dwarfs will slowly transform their appearance into a DZ spectrum, with no detectable Swan bands remaining \citep{blouin2022}.  First, it must be noted that roughly one quarter of the modeled DZQ stars in that study are likely detectable with S/N $>100$ spectroscopy, which is thus a promising avenue for further investigation.  Second, and perhaps more importantly, this hypothesis cannot account for the other properties of DQ stars; their lack of circumstellar debris, relatively low masses, lack of close companions, and the fact that their kinematics appear modestly distinct from those of DZ white dwarfs.  Hence while this possibility cannot be ruled out and is readily testable, it does not appear to tick most of the necessary boxes in Table~\ref{hypoths}.  Figure~\ref{vma2} demonstrates that a carbon abundance of [C/He] $=-6.5$ can be ruled out for vMa\,2, the DZ white dwarf prototype at $T_{\rm eff} = 6100$\,K \citep{coutu2019}.  While modestly smaller traces of carbon are possible at this effective temperature, corresponding to the cool end of the classical DQ white dwarf sequence, more stringent tests are possible around warmer DZ stars.

\subsection{Metals are masked by unknown stellar opacity}

This hypothesis states that the heavy element pollution in DQ white dwarfs is present, and similar to that observed e.g.\ in DZ stars, but is masked by an unknown stellar opacity source.  This possibility can in principle account for the fact that DQ stars rarely exhibit {\em detectable} metal lines, but does not address the bulk of relevant issues discussed in the previous section.  The derived masses of metals in the fully mixed outer layers of the star, as well as the inferred accretion rates, are tied to line strengths.  Then, if metal lines were reduced in strength within DQ stars and not in DZ stars, owing to a previously unaccounted for opacity source, there would be an expected trend with effective temperature or carbon abundance (or both).  Neither the deep UVES observations, nor the hundreds of stars observed with the SDSS, show any trends in detected Ca\,{\sc ii} line strengths with the DQ stellar parameters.  This hypothesis fails to address most of the other outstanding characteristics of the DQ stars, and would not, for example, prevent the detection of circumstellar material in the infrared, nor account for their distinctive stellar properties.

\subsection{Metals are diluted in deeper convective layers}

This idea is similar to that discussed above, but instead dilutes atmospheric metals within significantly deeper convection zones.  A scenario where the relative sizes of the fully mixed outer layers of DQ and DZ stars are orders of magnitude different can, in principle, account for the offset seen in the right hand panel of Figure~\ref{mdots}.  This could be achieved by increasing the size of convection zones in DQ stars, or decreasing those in DZ stars.  However, any order of magnitude changes in the depth of these layers will also directly effect the heavy element diffusion timescale, as these two are strongly correlated \citep{koester2009a}.  For example, if DQ convection zones were $1000\times$ larger than currently predicted, and crudely extrapolating linearly from current models, one might expect diluted metals in DQ stars to be seen for timescales of Gyr, and thus more frequently than in DZ stars, all else being equal.  Unfortunately, this hypothesis faces all the same problems as the previous, and would have to be reconciled with the current underlying framework of diffusion in helium atmosphere white dwarfs.

\subsection{Progenitors of DQ stars have high-intermediate masses}

A plausible theory for the lack of planetary material in and around DQ white dwarfs is inhibited planet formation, and one possibility is via higher-mass main-sequence stars.  If DQ white dwarfs are the progeny of stars with masses greater than $3-4$\,M$_{\odot}$, then it could be argued that this is consistent with their lack of circumstellar matter, and pollution from planetary debris.  Around such luminous stars, protoplanetary disk lifetimes may be insufficient to form planetary precursors and full-fledged planets \citep{kennedy2009,kunimoto2021}, and there are currently few confirmed cases of planets detected towards stars with masses above 3\,M$_{\odot}$ (e.g.\ \citealt{reffert2015}).  However, the expectations of this hypothesis would be that the remnant masses of DQ white dwarfs would be higher than the field, not lower as observed and modeled.  And while such progeny may be prone to detectable magnetism \citep{caiazzo2020}, they should be a relatively young kinematical population, having spent little time on the main sequence.  These predictions fail to match the properties of DQ stars as shown in Figure~\ref{chist}.  Furthermore, higher-mass parent stars would not address the lack of post-common envelope binaries, unless it were shown that this leads more commonly to companion destruction during the post-main sequence.

%%%TABLE HYPOTHS%%%
\begin{table}
\setlength{\tabcolsep}{5pt}
\begin{center}
\caption{Consistency of DQ hypotheses vs. observational constraints.\label{hypoths}}
\begin{tabular}{@{}ccccccc@{}}

\hline

Constraint			&Carbon	&Metal	&Metals 	&High 	&Metal	&Binary\\
				&Masked	&Masked	&Diluted	&Mass	&Poor	&Evol\\
\hline

Pollution			&+		&+		&+		&+		&+		&+\\
Circumstellar		&--		&--		&--		&+		&+		&+\\
Companions		&--		&--		&--		&--		&--		&+\\
$q({\rm He})$ / Mass	&+		&--		&--		&--		&+		&+\\
Magnetism		&--		&--		&--		&+		&--		&+\\
Kinematics		&--		&+		&+		&--		&--		&+\\

\hline

\end{tabular}
\end{center}
\end{table}

%%% 0913+103 COMPARISON%%%
\begin{figure}
\includegraphics[width=\columnwidth]{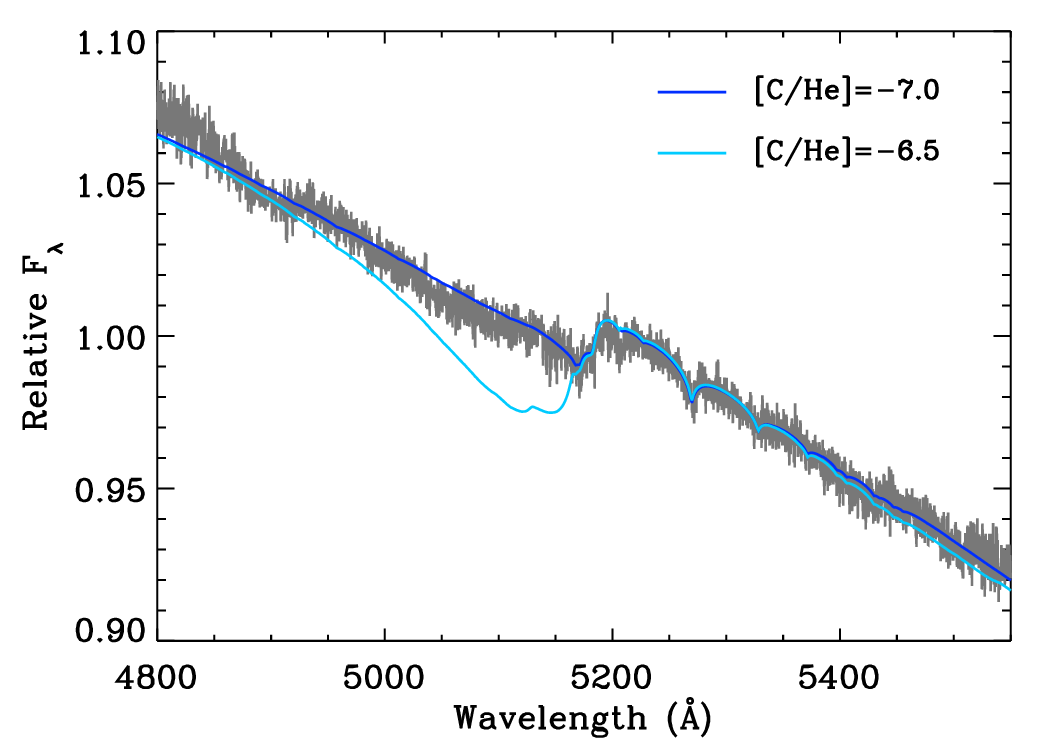}
\vskip 0pt
\caption{A VLT X-shooter  spectrum of the prototype DZ white dwarf vMa\,2 \citep{vanmaanen1917}, with S/N $\approx200$ in the displayed wavelength range (taken 2016 Nov 4 for program 098.D-0392).  Overplotted are two atmospheric models that include metal abundances informed by the strong lines present below 4000\,\AA, and the Mg\,{\sc i} 5183\,\AA \ feature seen in the plot, but with additional traces of carbon.  The model with [C/He] $=-6.5$ is readily ruled out, but would otherwise be among one of the lowest carbon abundances detected in any white dwarf \citep{koester2019,blouin2019}.
\label{vma2}}
\end{figure}

\subsection{Progenitors of DQ stars are (very) metal poor}

Similar to the hypothesis above, a lack of planetary matter orbiting and polluting DQ stars could be attributed to an origin as Population II stars that are truly metal poor.  There are numerous planetary occurrence rate studies that focus on stellar metallicity for hosts of transiting or Doppler planets, but in general these are restricted to metallicities higher than 0.1\,$Z_{\odot}$.  The most ancient planetary systems confirmed to date orbit thick disk stars that are only modestly metal poor \citep{mortier2012,campante2015}, and similarly debris disk masses are thought to be strongly linked to metallicities comparable to solar \citep{gaspar2016}.  Truly metal poor stars have not yet revealed planetary systems, and theoretical models suggest there is a critical metallicity, below which planet formation becomes challenging, with values in the range $0.01 < Z/Z_{\odot} < 0.1$ \citep{johnson2012,hasegawa2014}.  If DQ white dwarfs descend from such metal-poor stars, that could account for the lack of planetary material as pollution and circumstellar debris, and might also be consistent with their low masses, as the remains of stars that spent the bulk of their lifetime on the main sequence.  However, as shown above, their kinematics utterly fail to support this interpretation.  Furthermore, the known widely-bound companions to DQ stars are not significantly metal poor, and one would expect some fraction to host e.g.\ M subdwarf companions \citep{monteiro2006,zhang2013}.

\subsection{Progenitors of DQ stars are binaries}

A binary origin for DQ white dwarfs has several advantages over other hypotheses.  This scenario can immediately be invoked to account for the apparently low masses, and more acutely, their relatively thin helium layer mass fractions, which are arguably their defining characteristics according to current models \citep{coutu2019,koester2019,bedard2022}.  
Furthermore, it is already generally accepted that binary evolution is responsible for the warmer examples of DQ stars \citep{dunlap2015,cheng2020}, and while precise evolutionary modeling has yet to reproduce this inference based on observations, at face value there is a clear pathway to reduce $q({\rm He})$ in DQ star progenitors via binary interactions.

During close binary evolution, a post-main sequence precursor star will often lose mass via unstable Roche-lobe overflow, initiating a common envelope that shrinks the semimajor axis, and which is known to lead to lower remnant masses \citep{marsh1995,rebassa2011}.  This scenario relieves the tension between the relatively young space motions of the DQ stars, and the long single-star lifetimes implied by their low masses.  A high incidence of (strong) magnetism might also be expected for a binary population, as is observed for DQ white dwarfs.

Of all the theories discussed above, only a binary population might account for the apparent lack of unevolved, post-common envelope companions, and it requires that such companions are either consumed, or remain hidden.  The major challenge for this hypothesis is to show that binary evolution does not favor the formation or retention of planetary architectures that commonly pollute the surfaces of white dwarfs, and theoretical work is needed to investigate this aspect in particular.  It can be speculated that asymmetric mass loss, especially that which might occur during the cannibalization of low-mass stellar or substellar companions, might assist in the removal or destabilization of planetary material within the system.

\section{A DQ binary population outlook}

This study has shown that DQ white dwarfs originate in a stellar and circumstellar population that appear distinct from the DAZ, DBZ, and DZ hosts of remnant planetary systems, and highlighted several other, potentially disparate characteristics of these carbon-enriched stars that suggest a binary origin.  It is noteworthy that, for the most part, this work has focussed on the classical and cool DQ stars that are defined spectroscopically by the presence of Swan bands.  However, there is a miniature zoo of DQ spectral subtypes, and in particular a binary origin has already been suggested for those DQ stars that have warmer temperatures and even thinner, or absent, helium layers \citep{dunlap2015,cheng2020}, as well as unusual spectral types such as DAQ \citep{hollands2020}, all of which appear to have higher remnant masses, and thus superficially amenable to merger hypotheses.  These deductions have yet to be supported by dedicated evolutionary modeling efforts, but if correct, then one clear outcome is that binary interactions can result in reduced $q({\rm He})$ in white dwarf remnants.  

Therefore, if binary evolution can reduce the helium mass fraction in white dwarfs, the outlook is immediately positive for the entire class of DQ stars.  The only major difference for the cool and classical DQ stars with Swan bands, those that form the bulk of the known population, is their lower or more typical remnant masses.  For this reason, the merger of two white dwarfs is unlikely to be the dominant, binary formation channel for DQ stars.  Moreover, the classical DQ white dwarfs do not have kinematics of a blue straggler population, with an added delay in cooling for the time it takes to merge.  Binary evolution does not necessarily imply older ages, as two stars can begin to interact as soon as one has evolved sufficiently for mass transfer to begin, and thus a binary DQ population can have ages similar to their helium atmosphere, DZ and DC white dwarf counterparts, as observed.  The common envelope is thought to be a rapid process, and should not add significantly to the inferred total ages of DQ stars in the case that unevolved companions are immediately consumed; the age bottleneck is still the evolutionary timescale of the white dwarf progenitor.  

Planetary companions are also a possibility in a scenario of fatal engulfment and binary evolution for DQ white dwarfs.  Two basic outcomes are required, one is that the white dwarf experiences enhanced mass loss compared to single star evolution (possibly in terms of total mass and the outer helium envelope), and that any substellar companions prohibit or diminish planetary material that might later precipitate atmospheric pollution.  There is some evidence that giant planets in close orbits (i.e.\ hot Jupiters) are rarely found with additional planetary companions, and may have cleared away planet-forming disk material during migration, or ejected other planets dynamically \citep{latham2011,mustill2015}.  While it is unclear if a giant planet (or two) is sufficient to amplify mass loss during stellar evolution, the absence of hot Jupiters around subgiant hosts argues that giant planet ingestion does indeed occur \citep{schlaufman2013}.

A key question is exactly how binary evolution might prevent the eventual pollution of a white dwarf, where theoretical modeling is clearly needed.  From observations, it is noteworthy that for known white dwarfs in short-period binaries with low-mass stellar and substellar companions, there is only a single case of circumbinary material and pollution (SDSS\,J155720.77+091624.6; \citealt{farihi2017a}).  Despite this apparent rarity, there is a general lack of sensitivity to such circumbinary dust systems, because the cool companion can itself cause a large infrared excess, and thus additional emission from dust would be challenging to identify.

Binary evolution modeling is needed to test the basic premise raised here for DQ white dwarfs.  The lack of unevolved companions in close orbits is a key observable; on the one hand, if they are destroyed then the remnant might be expected to be somewhat more massive, and on the other hand if they survive, it is unclear why they are not detected.  It is theoretically possible for a companion to merge and eject more material than it provides to the merged stellar core, but such outcomes are best addressed via dedicated modeling.  It is important to note that both Procyon\,B and GJ\,86B may have transferred mass during their evolution, as the originally more massive stars within their binary systems.  Current data and modeling do not require such mass transfer in these iconic systems, but it may account for remaining inconsistencies, especially from an evolutionary perspective \citep{bond2015,zeng2022}.  Binary or white dwarf -- main-sequence merger candidates may be challenging to detect given the lack of atomic lines in many DQ stars, but relatively rapid stellar rotation, photometric or astrometric variation owing to unseen companions are likely one way forward, observationally.

Lastly, what is the nature of the rare and weak pollution in DQ white dwarfs?  The default hypothesis should be remnant planetary systems\footnote{Appealing to the interstellar medium for the weak DQZ pollution may be tempting, but there are myriad issues such as their utter lack of hydrogen.}, and there is no a priori reason to expect other mechanisms, even if the stars are currently or formerly binary \citep{rafikov2013,martin2013,bromley2015}.  There are dynamical issues that are unique to binary systems, and circumbinary planet hosts in particular, with additional instabilities as compared to single stars, during formation and on the main sequence, where planetary ejection is one of the likely outcomes \citep{smullen2016,sutherland2016,sutherland2019}.  It might be speculated that such instabilities might be catastrophically amplified during a common envelope or merger event where mass loss may be asymmetric or impulsive, thus quickly reducing or removing planets and associated planetesimal belts.  

For context, the solar zodiacal cloud is replenished at a rate in the range $10^7-10^8$\,g\,s$^{-1}$ \citep{nesvorny2011,rigley2022}, and the rare DQZ exhibit accretion rates $100\times$ lower, not to mention the typical DQ upper limits at least $1000\times$ lower, which approaches the accretion rate of extraterrestrial material onto the Earth \citep{love1993,yada2004}.  It may be possible to detect multiple heavy elements in the rare DQZ stars, and if successful, their ratios can be compared to the planetary abundances associated with DAZ, DBZ, and DZ white dwarfs.  While challenging owing to the low abundances and line strengths, it may provide an important clue to the nature of these stars, which are now, hopefully, less of an astrophysical enigma.

\section*{Acknowledgements}

The authors thank an anonymous reviewer for a careful reading of the manuscript.  J.~Farihi thanks D.~Koester and G.~Fontaine for key discussions over the years that helped to form the foundation of this work, E.~Dennihy for confirming the fact that no DQ star has a suspected infrared excess from circumstellar dust, A.~Rebassa-Mansergas for sharing his full white dwarf-main sequence binary catalog, and valuable exchanges with E.~B.~Bauer, R.~R.~Rafikov, and J.~J.~Eldridge.  Several colleagues provided feedback on an earlier version of the manuscript, including those mentioned above, as well as T.~von Hippel, A.~J.~Mustill, A.~Swan, and B.~Zuckerman.  The authors acknowledge the European Southern Observatory for the award of telescope time via programs 095.D-0706, 096.D-0076, and 097.D-0063.  This work has made use of the ESA {\em Gaia} mission, processed by the {\em Gaia} DPAC.  J.~Farihi  acknowledges support from STFC grant ST/R000476/1.  T.~G.~Wilson acknowledges support from STFC consolidated grants ST/R000824/1, ST/V000861/1, and UKSA grant ST/R003203/1.

\section*{Data Availability}
All spectra are available through public archives held at the European Souther Observatory or the Sloan Digital Sky Survey.

\bibliographystyle{mnras}

\bibliography{/Users/jfarihi/papers/references}

\bsp    % typesetting comment
\label{lastpage}
\end{document}